\documentclass[aps,twocolumn,superscriptaddress,nofootinbib,longbibliography]{revtex4-1}
\usepackage{booktabs}
\usepackage{amsmath}
\usepackage{amsthm}
\usepackage{amssymb}
\usepackage{esint}

\makeatletter

\providecommand{\tabularnewline}{\\}

 \usepackage{times}
\usepackage{graphicx}
\usepackage[section]{placeins}%%put table in its section
\usepackage{tabularx}
\usepackage{amsmath}
\usepackage{amstext}
\usepackage{amssymb}
\usepackage{xfrac}
\usepackage[colorlinks,citecolor=blue]{hyperref}
\usepackage{graphicx}
\usepackage{amsmath}
\usepackage{amstext}
\usepackage{amssymb}
\usepackage{amsfonts}
\usepackage{longtable,booktabs}
\usepackage{hyperref}
\usepackage{url}
\usepackage{subfigure}
\usepackage{dsfont}
\usepackage{booktabs}
\usepackage{amsbsy}
\usepackage{dcolumn}
\usepackage{amsthm}
\usepackage{bm}
\usepackage{esint}
\usepackage{multirow}
\usepackage{hyperref}
\usepackage{cleveref}
\usepackage{mathrsfs}
\usepackage{amsfonts}
\usepackage{amsbsy}
\usepackage{dcolumn}
\usepackage{bm}
\usepackage{multirow}
\usepackage{color}
\usepackage{extarrows}
\usepackage{datetime}
\usepackage{comment}
\usepackage[super]{nth}

\hypersetup{
    colorlinks=true,
    linkcolor=blue,
    filecolor=magenta,
        urlcolor=blue,
}

\newcommand{\comments}[1]{}

\def\Z{\mathbb{Z}}

\makeatother

\begin{document}
  \title{Continuum field theory of $3$D topological orders with emergent fermions and   braiding statistics}

 \author{Zhi-Feng Zhang}
 \affiliation{School of Physics, Sun Yat-sen University, Guangzhou 510275, China}
\author{Qing-Rui Wang}
\email{wangqr@mail.tsinghua.edu.cn}
\affiliation{Yau Mathematical Sciences Center, Tsinghua University, Haidian, Beijing 100084, China}
\author{Peng Ye}
\email{yepeng5@mail.sysu.edu.cn}
\affiliation{School of Physics, Sun Yat-sen University, Guangzhou 510275, China}
\date{\color{blue} {\today}}
\begin{abstract}
Universal topological data of topologically ordered phases can be captured by topological quantum field theory in continuous space time by taking the limit of low energies and long wavelengths. While previous continuum field-theoretical studies of topological orders in $3$D real space focus on either self-statistics, braiding statistics, shrinking rules, fusion rules or quantum dimensions, it is yet to systematically put all   topological data together in a unified continuum field-theoretical framework.  Here, we construct the topological $BF$ field theory   with   twisted terms (e.g., $AAdA$ and $AAB$) as well as a $K$-matrix $BB$ term, in order to simultaneously explore all such topological data and reach anomaly-free topological orders. Following the spirit of the famous $K$-matrix Chern-Simons theory of $2$D topological orders, we present general   formulas and systematically show how the $K$-matrix $BB$ term confines topological excitations, and how self-statistics of  particles is transmuted between bosonic one and fermionic one. In order to reach anomaly-free topological orders, we explore,  within  the present continuum field-theoretical framework, how the principle of gauge invariance fundamentally influences possible realizations of topological data. More concretely, we present the topological actions of (i) particle-loop braidings with emergent fermions, (ii) multiloop braidings with emergent fermions, and (iii) Borromean-Rings braidings with emergent fermions, and calculate their universal topological data.    Together with the previous efforts, our work paves the way toward a more systematic and complete continuum field-theoretical analysis of exotic topological properties of  $3$D topological orders. Several interesting future directions are also discussed.
\end{abstract} 
\maketitle

\tableofcontents{}

\section{Introduction}

Exploring low-energy long-wavelength effective field theories of quantum many-body systems has a long history in condensed matter physics~\cite{fradkin2013field}. For example, the Ginzburg-Landau (GL) field theory, in terms of local order parameters, is applied to symmetry-breaking phases and phase transitions; non-linear sigma models with topological $\theta$ term are applied to quantum spin chains. Since the discovery of the fractional quantum Hall effect in the 1980s, the notion of topological order has been introduced as a route toward exotic phases of matter that cannot be characterized by the mechanism of symmetry-breaking. While there has been a broad consensus that the essence of topological order is deeply rooted in patterns of long-range entanglement that is robust against local unitaries of finite depth~\cite{Chen10}, the original definition of \textit{topological} order really comes from the fact that the low energy effective field theory of the prototypical topological order--fractional quantum Hall states---is the Chern-Simons theory which is a \textit{topological} quantum field theory (TQFT)~\cite{Witten1989,Turaev2016} in continous spacetime. Along this line of thinking, the common expectation for topological phases of matter---topological robustness against any local perturbations---is achievable by simply noting the fact that correlation functions of all spatially-local operators in TQFTs vanish~\cite{sarma_08_TQC}, which is in sharp contrast with the GL theory. Particularly, as the most general Abelian formulation, the $K$-matrix Chern-Simons theory (KCS)~\cite{PhysRevB.42.8145,PhysRevB.46.2290} whose action is written in terms of $\sim \int \frac{K_{IJ}}{4\pi}A^IdA^J$, serves as the standard TQFT framework of $2$D Abelian topological orders, providing a highly efficient algorithm for computing topological data, such as anyon types, self-statistics, mutual statistics, fusion algebra, chiral central charge, and ground state degeneracy. Besides topological orders, the KCS has also been successfully applied to the study of symmetry-enriched topological phases (SET)~\cite{PhysRevB.93.155121,PhysRevB.87.195103} and symmetry-protected `topological' phases~\cite{PhysRevB.86.125119,YW12,PhysRevB.93.115136,Ye14b,PhysRevLett.112.141602} where global symmetry is nontrivially imposed. 

While the KCS works very well in $2$D topological phases of matter, it is no longer applicable to $3$D and higher where exotic spatially extended excitations (e.g., loops in $3$D and higher, membranes in $4$D and higher) induce very rich   emergent phenomena. Instead, if particles and loops respectively carry gauge charges and gauge fluxes of a discrete Abelian gauge group $G=\prod_{i=1}^n{\mathbb{Z}_{N_i}}$, one may apply the twisted ``$BF$ field theory''~\cite{Horowitz1990,hansson2004superconductors} by properly including ``twisted  terms'' (denoted as $T$)~\cite{2016arXiv161209298P,PhysRevB.99.235137,YeGu2015,ypdw,yp18prl}. This series of TQFTs have been proved to a very powerful way to efficiently describe various types of nontrivial braiding statistics, such as,  
   particle-loop braiding~\cite{hansson2004superconductors,abeffect,PRESKILL199050,PhysRevLett.62.1071,PhysRevLett.62.1221,ALFORD1992251}, multi-loop braiding~\cite{wang_levin1} (with the twisted terms $\sim AAdA$ and $\sim AAAA$), and particle-loop-loop braiding (i.e., Borromean-Rings braiding with the twisted term $\sim AAB$) \cite{yp18prl}.  Recently, the untwisted / twisted $BF$ theory has also been successfully applied to SET~\cite{Ning2018prb,ye16_set,2016arXiv161008645Y,Ye:2017aa,ye16a} and SPT~\cite{YeGu2015,bti2,PhysRevLett.114.031601,PhysRevB.99.205120,bti6} in $3$D.

In the twisted $BF$ theory (symbolically denoted as ``$BF+T$''),  each type of braiding statistics is associated with a particular formulation of TQFT actions, which does not mean that all types of braiding statistics are mutually compatible and can thereby coexist in an anomaly-free topological order.   To further examine whether two different types of braiding processes are allowed to compatibly exist in the same topological order, Ref.~\cite{zhang2021compatible} exhausted all combinations of twisted terms and found that TQFTs of some combinations inevitably violate the principle of gauge invariance. Thus, among all combinations, only a part of combinations are legitimate such that braiding processes can coexist. After TQFTs with mutually compatible braiding processes are obtained,   fusion rules and shrinking rules in the TQFTs are further investigated, from which quantum dimensions of both particles and loops are computed~\cite{PhysRevB.107.165117}. Recently, the ideas of Ref.~\cite{zhang2021compatible} and Ref.~\cite{PhysRevB.107.165117} have been subsequently extended to $4$D real space \cite{Zhang:2021ycl,Huang:2023kfv} where membrane excitations are allowed and hierarchy of shrinking rules is definable.

On the other hand, in the untwisted $BF$ theory with the inclusion of a $BB$ term~\citep{horowitz89} (symbolically denoted as ``$BF+BB$''),   the boson-fermion statistical transmutation of self-statistics (i.e., exchange statistics) of particles in $3$D~\cite{PhysRevB.99.235137} has been studied through equations of motion, where the scenario of Dirac-string-attachment implied by the equations of motion mimics, to some extent, the physics of dyons studied intensively in other contexts~\cite{YW13a,witten1,ye16a,goldhaber89,PhysRevLett.36.1122}. Along this line, it has become clear that all particles with anyonic statistics (neither fermionic nor bosonic) are exactly confined and thus disappear in the low energy spectrum, which is perfectly consistent with the well-known fact that anyons are impossible in $3$D and higher~\cite{Leinaas1977,wu84,wilczek1990fractional}. Thanks to the statistical transmutation induced by the $BB$ term, we may realize both \textit{emergent fermions} (defined as topologically nontrivial particles that are fermionic) and transparent fermions (defined as topologically trivial particles that are fermionic) in a topological action that is composed of  merely bosonic degrees of freedom (i.e., gauge fields). As a side, by definition, once transparent particles are fermionic,  the topological order is said to be  \textit{fermionic}.
In addition to boson-fermion transmutation, the single-component $BB$ term also provides a novel ``Higgs'' mechanism that confines either partially or completely the gauge group $G$ set by the coefficient of the $BF$ term~\cite{Kapustin:2014gua}.  Besides, the multi-component $BB$ term was   successfully applied to $3$D bosonic
topological insulators (bosonic SPTs with particle number conservation and time-reversal symmetry) where bulk topological order is trivial but boundary admits anomalous surface topological orders~\citep{bti2}.

Logically, once we have understood \textbf{(\textit{i})} how to obtain compatible braiding processes via legitimate combinations of twisted terms in the twisted $BF$ theory ($BF+T$) and \textbf{(\textit{ii})} how to assign self-statistics on particles via the boson-fermion transmutation in the untwisted $BF$ theory with the $BB$ term ($BF+BB$), it becomes urgent to make a step forward by examining whether braiding statistics is compatible with the assignment of self-statistics on particles in the twisted $BF$ theory with the $BB$ term  (denoted as ``$BF+T+BB$'') \textit{within the present continuum-field-theoretical framework}. We are motivated to  combine all known topological terms in continuous spacetime in order to  achieve a more complete continuum-field-theoretical description of topological data encoded in $3$D topological orders with the gauge group $G$.

In this paper, we first explain the microscopic origin of topological terms via various condensation pictures at the beginning of  Sec.~\ref{section_KBB}, in order to \textit{(i)} make the gauge theories more physical in the context of many-body physics and \textit{(ii)} introduce the critical role of Lagrange multipliers. The topological action of a  $\mathbb{Z}_{N}$ topological order is dual to a standard Abel-Higgs model that describes a boson/vortex-line condensate coupled to gauge field. Other topological twisted terms and $B  B$ term can be formally derived through introducing topological interactions among different condensates.  In the remaining part of Sec.~\ref{section_KBB}, we systematically formulate the untwisted $BF$ theory with a $K$-matrix $BB$ term (i.e., $\sim\frac{K_{ij}}{4\pi}B^iB^j$ with a symmetric integer matrix $K$). In the presence of the $K$-matrix $BB$ term, we present general mathematical formulas that can be applied to efficiently determine (\textit{i}) excitation contents, i.e., inequivalent Wilson operators of deconfined particles and deconfined loops (Fig.~\ref{fig_higgs}), and (\textit{ii}) self-statistics assignment on particles.  In particular, the situation of the single-component $BB$ term has been naturally included by regarding the $K$ matrix as an integer. Then, the self-statistics of particles is rigorously derived by computing the expectation values of \textit{framed} Wilson loops (Fig.~\ref{fig_framing}). We obtain the formula (\ref{eq_general_formula_self}) in a compact form that completely fixes self-statistics of the particle labeled by an integer vector, whose usefulness is comparable to the familiar formula in the famous KCS theory of $2$D topological orders~\cite{PhysRevB.42.8145,PhysRevB.46.2290}. In Table~\ref{table_singleBB_properties}, we collect, for our purpose, the most useful properties of the untwisted $BF$ theory with a single-component $BB$ term. To determine whether a topological order is fermionic or bosonic, we may calculate the self-statistics of trivial (i.e., transparent) particles. To determine whether a topological order supports emergent fermions, we may calculate the self-statistics of particles that carry nontrivial gauge charges of $G$.

  Sec.~\ref{section_compatible}  is devoted to studying the interplay of  topological data including self-statistics, braiding statistics as well as fusion rules. For this purpose,  $BF$ theories with $BB$ term and different twists are studied, leading to ``$BF+T+BB$''. Three root braiding processes (coined in Ref.~\cite{zhang2021compatible}) and their braiding phases are considered: particle-loop braiding ($BF$ term), multi-loop braiding ($AAdA$ and $AAAA$ twist), and Borromean rings braiding ($AAB$ twist).  For completeness, we start our discussions in Sec.~\ref{sec_plbraid} by considering particle-loop braiding with emergent fermions, which is described by  $BF+BB$.   Then, we move to continuum field theory description of coexistence of emergent fermions and multi-loop braidings (Sec.~\ref{sec_mlbraid}) and coexistence of emergent fermions and Borromean-Rings braiding (Sec.~\ref{sec_brbraid}). We present topological actions, gauge transformations for each case, and calculate topological data including inequivalent Wilson operators, braiding statistics, self-statistics, fusion rules, shrinking rules, and quantum dimensions.  We draw our conclusions and make some discussions in Sec.~\ref{section_conclusion}.

\section{Condensation picture and $K$-matrix $BB$ term }\label{section_KBB}

 \subsection{Condensation picture via Abel-Higgs models with topological interactions}\label{subsec_condensation_picture}
Here we review a condensation
picture for TQFT in $\left(3+1\right)$D \citep{hansson2004superconductors,PhysRevB.93.115136,bti2,YeGu2015,zhang2021compatible}.
A $\mathbb{Z}_{N}$ topological order described by a topological action
$S=\int\frac{N}{2\pi}B\wedge dA$ can be viewed as the Higgs phase
of a Abel-Higgs model. Such a model describes a condensate of charge
boson (or flux-threaded vortex-lines) coupled to a gauge field. In this
section, we will explain this microscopic origin of topological actions.
We start from a single layer condensate of boson or vortex-line. Then,
by turning on topological interactions among different condensates, other topological
terms emerge after a duality transformation.

We first show how to derive the topological $BF$ term from a boson
condensation (or a vortex-line condensation) coupled to a gauge field.
Consider a condensation of charge-$N$ bosons that couple to a gauge
field: $\mathcal{L}=\frac{\rho}{2}\left(\partial_{\mu}\theta-NA_{\mu}\right)^{2}+\mathcal{L}_{\textrm{Maxwell}}$
where $A$ is a $U\left(1\right)$ gauge field. This is nothing  
but the deconfined phase of the Abel-Higgs model. With a Hubbard-Stratonovich auxiliary
field $j_{\mu}$, this Lagrangian is dual to $\mathcal{L}_{d}=-\frac{1}{2\rho}\left(j_{\mu}\right)^{2}+j_{\mu}\left(\partial_{\mu}\theta-NA_{\mu}\right)$.
Integration of $\theta$ results in a constraint $\delta\left(\partial_{\mu}j_{\mu}\right)$
in the path integral measure which can be resolved by introducing
a $2$-form gauge field $B_{\mu\nu}$: $j_{\mu}=-\frac{1}{4\pi}\epsilon^{\mu\nu\lambda\rho}\partial_{\nu}B_{\lambda\rho}$.
Substituting this solution to the dual Lagrangian and dropping the
irrelevant Maxwell terms, we obtain $\mathcal{L}_{d}\sim\frac{N}{4\pi}\epsilon^{\mu\nu\lambda\rho}A_{\mu}\partial_{\nu}B_{\lambda\rho}$.
Then the action is $S=\int\mathcal{L}_{d}dxdt\sim\int\frac{N}{2\pi}A\wedge dB$
where$A=\sum_{\mu}A_{\mu}dx^{\mu}$ and $B=\frac{1}{2!}\sum_{\mu\nu}B_{\mu\nu}dx^{\mu}dx^{\nu}$.
Through integration by parts and dropping the total derivative term,
we reach the topological $BF$ term, $S=\int\frac{N}{2\pi}B\wedge dA$.
In this action, $B$ serves as a Lagrange multiplier to enforce $dA=0$
locally. 

On the other hand, we can also consider a condensation of flux-$N$
vortex-lines \citep{bti2} coupled to a $2$-form gauge field: $\mathcal{L}'=\frac{\rho}{2}\left(\partial_{[\mu}\Theta_{\nu}]-NB_{\mu\nu}\right)^{2}+\mathcal{L}'_{\textrm{Maxwell}}$
where $\Theta$ is the phase of vortex-line condensation, $B$ is
a $U\left(1\right)$ gauge field, and $\partial_{[\mu}\Theta_{\nu]}=\partial_{\mu}\Theta_{\nu}-\partial_{\nu}\Theta_{\mu}$.
This actually is another kind of Abel-Higgs model of a $2$-form gauge
field. Similar to previous discussion, $\mathcal{L}'$ is dual to
$\mathcal{L}'_{d}=-\frac{1}{8\rho}\left(\Sigma_{\mu\nu}\right)^{2}+\frac{1}{2}\Sigma_{\mu\nu}\left(\partial_{[\mu}\Theta_{\nu}]-NB_{\mu\nu}\right)$
with a Hubbard-Stratonovich auxiliary field $\Sigma_{\mu\nu}$. Integrating
over $\Theta$ leads to a constraint $\partial_{\nu}\Sigma_{\mu\nu}=0$
which can be resolved by $\Sigma_{\mu\nu}=-\frac{1}{2\pi}\epsilon^{\mu\nu\lambda\rho}\partial_{\lambda}A_{\rho}$.
Once again we   arrive at $\mathcal{L}'_{d}\sim\frac{N}{4\pi}\epsilon^{\mu\nu\lambda\rho}B_{\mu\nu}\partial_{\lambda}A_{\rho}$
and $S'=\int\mathcal{L}_{d}'dxdt=\int\frac{N}{2\pi}A\wedge dB+\cdots$,
where $\cdots$ includes Maxwell terms and boundary term that can
be dropped. In this case, it is $A$ that plays the role of Lagrange
multiplier to enforce $dB=0$ locally. $A\wedge dB$ is also a $BF$
term since it differs from $B\wedge dA$ by a total derivative. 

From the above discussion, we have seen that there are two kinds of Abel-Higgs
models that can be dual to the $BF$ term. The first (second) one
realizes the Higgs phase of a $1$-form ($2$-form) gauge theory. This also reveals that a $\mathbb{Z}_{N}$
topological order has a condensation picture: it originates from either
a boson condensate coupled to gauge field or a vortex-line condensate
coupled to gauge field. The $\mathbb{Z}_{N}$ gauge group structure
is encoded in the value of Wilson operator of $1$-form gauge field
or $2$-form gauge field. This picture can be generalized to a $\prod_{i=1}^{n}\mathbb{Z}_{N_{i}}$
topological order. We can derive different topological terms, e.g.,
$AAdA$, $AAAA$, and $AAB$ ($\wedge$ is omitted), through topological interactions among
 different condensates. 

For a $AAdA$ type topological term, its microscopic origin can be
traced back to a two-layer or three-layer condensates of charged bosons
\citep{YeGu2015}, e.g., $\mathcal{L}=\sum_{i=1}^{3}\frac{\rho_{i}}{2}\left(\partial_{\mu}\theta^{i}-N_{i}A_{\mu}^{i}\right)^{2}+{\rm i}q\epsilon^{\mu\nu\lambda\rho}\left(\partial_{\mu}\theta^{1}-N_{1}A_{\mu}^{1}\right)\left(\partial_{\nu}\theta^{2}-N_{2}A_{\nu}^{2}\right)\partial_{\lambda}A_{\rho}^{3}+\mathcal{L}_{\textrm{Maxwell}}$
where $q$ is a proper coefficient. The theory is dual to $S\sim\int\sum_{i=1}^{3}\frac{N_{i}}{2\pi}B^{i}dA^{i}+qA^{1}A^{2}dA^{3}$
that captures the three-loop braiding. For a $AAAA$ type topological
term, it can be derived from a four-layer condensate where each layer
is in charge-$N_{i}$ \emph{boson }condensation \citep{PhysRevB.93.115136},
e.g., $\mathcal{L}=\sum_{i=1}^{4}\frac{\rho_{i}}{2}\left(\partial_{\mu}\theta^{i}-N_{i}A_{\mu}^{i}\right)^{2}+{\rm i}q\epsilon^{\mu\nu\lambda\rho}\times\left(\partial_{\mu}\theta^{1}-N_{1}A_{\mu}^{1}\right)\times\left(\partial_{\nu}\theta^{2}-N_{2}A_{\nu}^{2}\right)\times\left(\partial_{\lambda}\theta^{3}-N_{3}A_{\lambda}^{3}\right)\times\left(\partial_{\rho}\theta^{2}-N_{4}A_{\rho}^{4}\right)+\mathcal{L}_{\textrm{Maxwell}}.$
This theory is dual to $S\sim\int\sum_{i=1}^{4}\frac{N_{i}}{2\pi}B^{i}dA^{i}+qA^{1}A^{2}A^{3}A^{4}$
that corresponds the four-loop braiding. Such condensation picture
applies for $AAB$ and $BB$ topological term with the caveat that
the $2$-form gauge field $B$ indicates a \emph{vortex-line} condensation.
The topological action $S=\int\sum_{i=1}^{3}\frac{N_{i}}{2\pi}B^{i}dA^{i}+qA^{1}A^{2}B^{3}$
can be derived from \citep{zhang2021compatible}: $\mathcal{L}=\sum_{i=1}^{2}\frac{\rho_{i}}{2}\left(\partial_{\mu}\theta^{i}-N_{i}A_{\mu}^{i}\right)^{2}+\frac{\left(\phi_{3}\right)^{2}}{2}\left(\partial_{[\mu}\Theta_{\nu]}^{3}-N_{3}B_{\mu\nu}^{3}\right)^{2}+{\rm i}q\left(\partial_{\mu}\theta^{1}-N_{1}A_{\mu}^{1}\right)\left(\partial_{\nu}\theta^{2}-N_{2}A_{\nu}^{2}\right)\times\left(\partial_{[\mu}\Theta_{\nu]}^{3}-N_{3}B_{\mu\nu}^{3}\right)+\mathcal{L}_{\textrm{Maxwell}}$
where layer $1$ and $2$ are in charge-$N_{1}$ and $N_{2}$ \emph{boson}
condensation while layer $3$ is in flux-$N_{3}$ \emph{vortex-line}
condensation. For the topological action $S=\int\frac{N_{1}}{2\pi}B^{i}dA^{i}+\frac{K_{11}}{4\pi}B^{1}B^{1}$
where $K_{11}$ is a proper coefficient, it can be derived from a condensate
of flux-$N_{1}$ vortex-line coupled to gauge field \citep{bti2}:
$\mathcal{L}=\frac{\phi^{2}}{2}\left(\partial_{[\mu}\Theta_{\nu]}^{s}-N_{1}B_{\mu\nu}^{1}\right)^{2}+{\rm i}K_{11}\epsilon^{\mu\nu\lambda\rho}\left(\partial_{[\mu}\Theta_{\nu]}^{s}-N_{1}B_{\mu\nu}^{1}\right)\left(\partial_{[\lambda}\Theta_{\rho]}^{s}-N_{1}B_{\lambda\rho}^{1}\right)+\mathcal{L}_{\textrm{Maxwell}}$.

Keeping this condensation picture in mind, we can examine these TQFT
actions more carefully. In order to describe a multi-loop braiding,
one can utilize $S\sim\int BdA+AAdA$ (three-loop braiding) or $S\sim\int BdA+AAAA$
(four-loop braiding). In these two actions, the $2$-form gauge field
$B$'s serve as Lagrange multipliers to enforce $dA=0$ locally. A
Borromean rings braiding can be described by $S\sim\int\sum_{i=1}^{3}\frac{N_{i}}{2\pi}B^idA^i+A^{1}A^{2}B^{3}$.
From the above derivation, we notice that $B^{1}$ and $B^{2}$ serve
as Lagrange multipliers while $A^{3}$ is the Lagrange multiplier
for layer $3$. Similarly, for the TQFT action $S\sim\int BdA+BB$,
$1$-form gauge field $A$ serves as Lagrange multiplier. We shall
emphasize the importance of Lagrange multiplier here. When we consider
an Abel-Higgs model of a one-form gauge field ($A$), a two-form gauge
field $B$ emerge as a Lagrange multiplier to encode the constraint
$dA=0$ in the path integral and vice versa. In other words, in a
$BF$ term $B\wedge dA$, either $A$ or $B$ serves as the Lagrange
multiplier, meaning that its microscopic origin can be derived from either a vortex-line condensation
or a boson condensation. When we discuss a $\prod_{i=1}^{n}\mathbb{Z}_{N_{i}}$
topological order, the $BF$ term is $\sum_{i=1}^{n}\frac{N_{i}}{2\pi}B^{i}dA^{i}$.
For each index $i$, only one of $A^{i}$ and $B^{i}$ is Lagrange
multiplier and there must be one Lagrangian multiplier such that a
$\mathbb{Z}_{N_{i}}$ gauge theory can be realized  in continuous spacetime. We can draw a conclusion that in our framework of continuum
field theory $S=S_{BF}+S_{{\rm int}}$, $A^{i}$ and $B^{i}$ cannot
be simultaneously involved in the interaction term $S_{{\rm int}}$,
i.e., $A^{i}$ and $B^{i}$ cannot show up in twisted terms and $BB$
term at the same time.

\subsection{$K$-matrix $BB$ term: topological action, gauge transformations, coefficient quantization and periods}
The untwisted $BF$ theory with a $K$-matrix $BB$ term is ($\wedge$ is omitted)
\begin{equation}
	S=\int\sum_{i=1}^{n}\frac{N_{i}}{2\pi}B^{i}dA^{i}+\sum_{i,j=1}^{n}\frac{K_{ij}}{4\pi}B^{i}B^{j}\,,\label{eq:n_component_BB term}
\end{equation}
where $K$ is an $n\times n$ symmetric matrix ($K_{ij}=K_{ji}$) whose quantization and periods will be determined shortly. The coefficients $\{N_i\}$ of the first term, i.e., the $BF$ term, determine the gauge group    $G=\prod_{i=1}^{n}\mathbb{Z}_{N_{i}}$.  The $[U(1)]^n\times [U(1)]^n$ gauge transformations of this gauge theory are defined as:  
\begin{align}
	A^{i}\rightarrow & A^{i}+d\chi^{i}-\sum_{j=1}^{n}\frac{K_{ij}}{N_{i}}V^{j}\,, \,\,\,B^{i}\rightarrow  B^{i}+dV^{i}\,,
\end{align}
where $\chi^i$ and $V^i$ are respectively $0$-form and $1$-form gauge parameters that satisfy the usual compactness conditions: $\frac{1}{2\pi}\int d \chi^i\in \Z$ and $\frac{1}{2\pi}\int d V^i\in \Z$. It is clear that the $BB$ term, denoted as $S_{BB}=\sum_{i,j=1}^{n}\frac{K_{ij}}{4\pi}B^{i}B^{j}$, induces an extra term ``$\sum_{j=1}^{n}\frac{K_{ij}}{N_{i}}V^{j}$'' compared to   the usual gauge transformations  of the $1$-form gauge field $A^i$.

 After    the  transformations, two additional terms are induced in the $BB$ term: $S_{BB}\rightarrow S_{BB}'=   S_{BB}+\Delta S_{BB}^{\left(1\right)}+\Delta S_{BB}^{\left(2\right)}\,$,
where
 $	\Delta S_{BB}^{\left(1\right)}= 2\int\sum_{i,j=1}^{n}\frac{K_{ij}}{4\pi}B^{i}dV^{j}$ and 
$	\Delta S_{BB}^{\left(2\right)}=  \int\sum_{i,j=1}^{n}\frac{K_{ij}}{4\pi}dV^{i}dV^{j}$. 
In a compact manifold, these two terms vanish if gauge parameters are topologically trivial. But in general, $\int d V^i$ can be nonzero. 
Here $\Delta S_{BB}^{\left(1\right)}$ can be written as $\Delta S_{BB}^{\left(1\right)}=2\int\sum_{i=1}^{n}\frac{K_{ii}}{4\pi}B^{i}dV^{i}+2\int\sum_{i<j}^{n}\frac{K_{ij}}{4\pi}B^{i}dV^{j}+2\int\sum_{i<j}^{n}\frac{K_{ji}}{4\pi}B^{j}dV^{i}$,
in which $\frac{1}{2\pi}\int B^{i}dV^{j}\in\frac{2\pi}{N_{i}}\mathbb{Z}$
for arbitrary $i$ and $j$. Demanding $\Delta S_{BB}^{\left(1\right)}\in2\pi\mathbb{Z}$,
we find constraints $\frac{K_{ii}}{N_{i}}\in\mathbb{Z}$, $\frac{K_{ij}}{N_{i}}\in\mathbb{Z}$ ($i<j$),
and $\frac{K_{ji}}{N_{j}}\in\mathbb{Z}$ ($i<j$). Recall $K_{ij}=K_{ji}$
and we find $\frac{K_{ij}}{N_{i}}\in\mathbb{Z}$ and $\frac{K_{ij}}{N_{j}}\in\mathbb{Z}$
for $i\neq j$. In fact, the constraint on $K_{ij}$ is $\frac{K_{ij}}{{\rm lcm}\left(N_{i},N_{j}\right)}\in\mathbb{Z}$
where ${\rm lcm}\left(N_{i},N_{j}\right)$ is the least common multiplier
of $N_{i}$ and $N_{j}$.

For the calculation of $\Delta S_{BB}^{\left(2\right)}$, we need
to consider whether a spin structure is taken into account. On a non-spin
manifold, $\frac{1}{4\pi^{2}}\int dV^{i}dV^{i}$ is quantized to $\mathbb{Z}$;
while on a spin manifold, it is quantized to $2\mathbb{Z}$. For $\frac{1}{4\pi^{2}}\int dV^{i}dV^{j}$
with $i\neq j$, it is quantized to $\mathbb{Z}$ no matter on a spin
or non-spin manifold. In order to keep $\Delta S_{BB}^{\left(2\right)}\in2\pi\mathbb{Z}$
for gauge invariance, we have  (\textit{i}) non-spin manifold: $K_{ii}\in2\mathbb{Z},K_{ij}\in\mathbb{Z}\, (i\neq j)$ and  (\textit{ii}) spin manifold: $ K_{ii}\in\mathbb{Z},K_{ij}\in\mathbb{Z} \,(i\neq j)$. 
Only on a spin manifold \textit{can} the diagonal elements $K_{ii}$  be an
odd integer. Indeed, as shown in the following main text, the parity of $K_{ii}$
controls the self-statistics of trivial particle excitations of
$\mathbb{Z}_{N_{i}}$ gauge subgroup. As long as one of the diagonal
elements $K_{ii}$ is odd, there must exist a fermionic trivial particle
excitation thus by definition the theory (\ref{eq:n_component_BB term}) describes 
a fermionic topological order. This is consistent with the fact that a fermionic
theory can only be defined on a spin manifold. On the other hand,
when all $K_{ii}$'s are even, this theory (\ref{eq:n_component_BB term}) is a bosonic one.

For the period of $K_{ij}$, we consider
 $	S_{BB}\in  \sum_{i=1}^{n}\frac{K_{ii}}{4\pi}\frac{\left(2\pi\right)^{2}}{N_{i}N_{i}}\mathbb{Z}+\sum_{i< j}^{n}2\cdot\frac{K_{ij}}{4\pi}\frac{\left(2\pi\right)^{2}}{N_{i}N_{j}}\mathbb{Z}\, 
 $. Since $\exp\left({\rm i}S_{BB}\right)$ should be invariant if we shift
either $\frac{K_{ii}}{4\pi}\frac{\left(2\pi\right)^{2}}{N_{i}N_{i}}$ or
$2\cdot\frac{K_{ij}}{4\pi}\frac{\left(2\pi\right)^{2}}{N_{i}N_{j}}$
by $2\pi$,  we have the following relations (we use $\simeq$
to denote such identification relation): 
 $	\frac{K_{ii}}{4\pi}\frac{\left(2\pi\right)^{2}}{N_{i}N_{i}} \simeq\frac{K_{ii}}{4\pi}\frac{\left(2\pi\right)^{2}}{N_{i}N_{i}}+2\pi
$ and 	$2\cdot\frac{K_{ij}}{4\pi}\frac{\left(2\pi\right)^{2}}{N_{i}N_{j}}  \simeq2\cdot\frac{K_{ij}}{4\pi}\frac{\left(2\pi\right)^{2}}{N_{i}N_{j}}+2\pi$. 
Therefore,  $	K_{ii}  \simeq K_{ii}+2\left(N_{i}\right)^{2}$ and $ K_{ij}  \simeq K_{ij}+N_{i}N_{j},i<j $.
 The period for $K_{ij}$ ($i>j$) is the same as that of $i<j$.

In conclusion,  the matrix elements of the symmetric $K$ matrix are simultaneously constrained by the following conditions (TO stands for topological order): 
\begin{align}
	& \frac{K_{ij}}{N_{i}}\in\mathbb{Z}\,,\frac{K_{ij}}{N_{j}}\in\mathbb{Z}\, (\forall i,j);\\
	& K_{ij}\in\mathbb{Z}\, (i\neq j);\\
	& K_{ii}\simeq K_{ii}+2\left(N_{i}\right)^{2};\\
	& K_{ij}\simeq K_{ij}+N_{i}N_{j}\,(i\neq j);\\
&	\textrm{For a bosonic TO: }  K_{ii}\in2\mathbb{Z};\\
&	\textrm{For a fermionic TO: at least one of \ensuremath{K_{ii}}'s is odd}.
\end{align}

\subsection{Wilson operators and (partial) confinement of gauge group}
\begin{figure}
	\centering
	\includegraphics[width=0.47\textwidth]{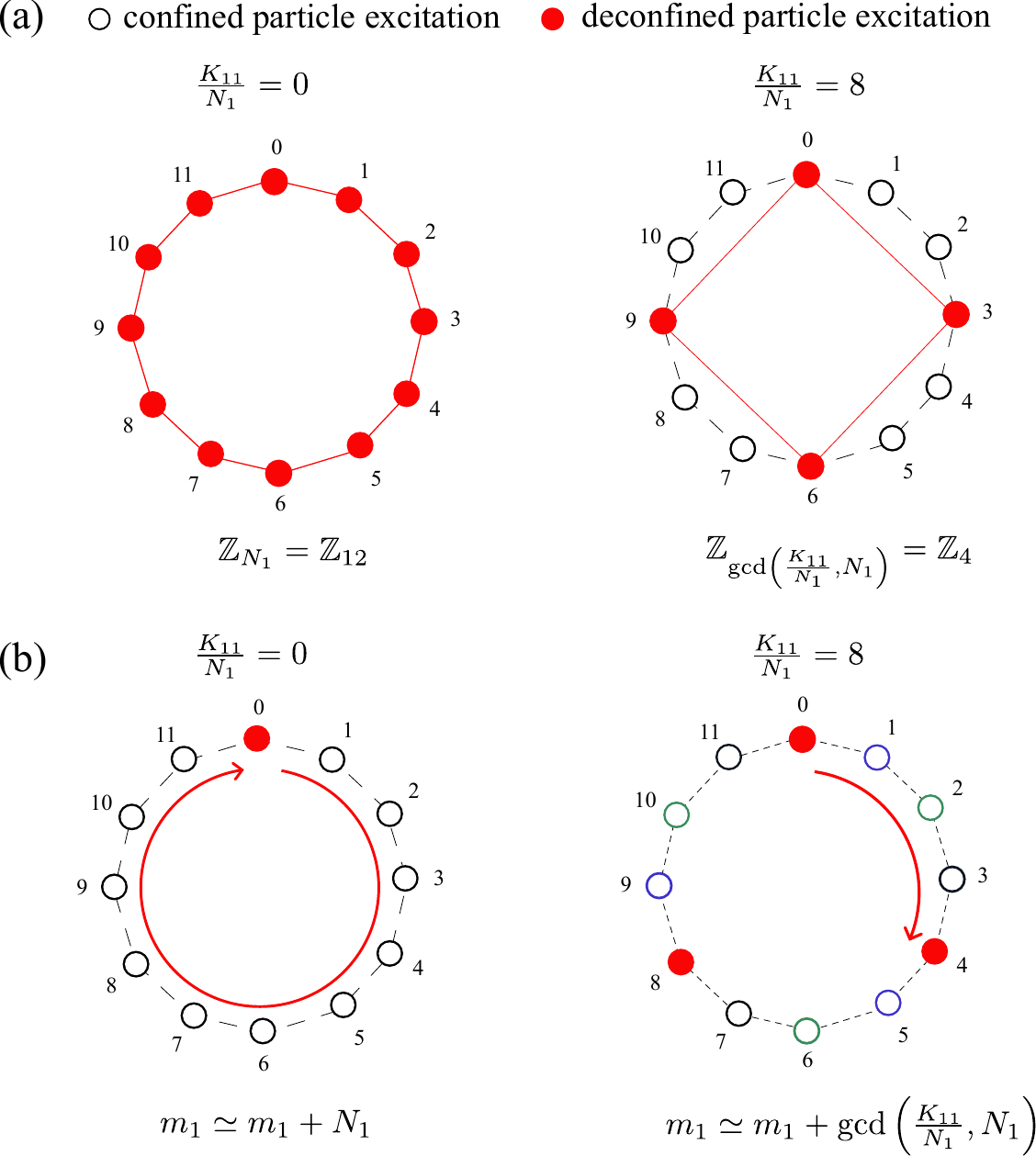}
	\caption{(a) When $\frac{K_{11}}{N_1}=0$, the charges of deconfined particle excitations of action (\ref{eq_action_BF+single_BB}) are labeled by $\Z_{N_1}=\Z_{12}$. Once a $B \wedge B$ term with $\frac{K_{11}}{N_1}$ is added, some particle excitations become confined. Those deconfined are labeled by the unbroken gauge group $\Z_{\gcd\left(\frac{K_{11}}{N_{1}},N_{1}\right)}=\Z_{4}$. (b) Period of fluxes for different values of $\frac{K_{11}}{N_1}$. When $\frac{K_{11}}{N_1}=0$, fluxes has a period of $N_1$. When $\frac{K_{11}}{N_1}=8$, some fluxes are actually equivalent as shown in the same color. In this case the minimal period of flux is $\gcd\left(\frac{K_{11}}{N_{1}},N_{1}\right)=4$. }\label{fig_higgs}
\end{figure}

A particle excitation carrying $e_i$ units of $\Z_{N_i}$ gauge charges can be labeled by a particle vector $\mathbf{l}=\left(e_{1},e_{2},\cdots,e_{n}\right)^{T}$
with $e_{i}\in \Z_{N_i}$, whose Wilson operator is
\begin{align}
	\!\!\!W\left(\mathbf{l},\gamma\right) & =\exp\left(\int_{\gamma}{\rm i}\sum_{i=1}^{n}e_{i}A^{i}+\sum_{i,j=1}^{n}\frac{{\rm i}e_{i}K_{ij}}{N_{i}}\int_{\Sigma_{i}^{j}}B^{j}\right)
	\label{eq:op_particle_n_component}
\end{align}
where $\Sigma_{i}^{j}$'s are Seifert surfaces of $\gamma$. The physical picture of (\ref{eq:op_particle_n_component}) is a particle excitation being attached by flux strings, see Fig.~\ref{fig_framing}(a). The amounts and species of fluxes are controlled by $K_{ij}$, elements of the $K$-matrix. Seifert surfaces $\Sigma_{i}^{j}$ with different $i,j$ correspond to the world sheets swap by different flux strings. Due to the tension on strings, a particle excitation may be confined. Only those attached by $2\pi$ fluxes are deconfined, i.e., 
 $	\frac{e_i K_{ij}}{N_i}\int_{\Sigma_{i}^{j}} B^j=\frac{e_i K_{ij}}{N_i}\frac{2\pi n_j}{N_j} \in 2\pi \Z 
 $,  
where $n_j$ is an integer. If a particle excitation labeled by $\mathbf{l}$ is deconfined, it is required that 
 $	\frac{e_i K_{ij}}{N_i N_j} \in \Z, \forall i,j \in \left\{1,\cdots,n\right\}.
$
For example, the constraints on $e_1$ are
 $	\frac{e_1 K_{11}}{N_1 N_1} \in \Z, \frac{e_1 K_{12}}{N_1 N_2} \in \Z, \cdots, \frac{e_1 K_{1n}}{N_1 N_n} \in \Z, 
 $ 
which demands	 $e_1 \in \frac{N_{1}}{\gcd\left(\frac{K_{11}}{N_{1}},\frac{K_{12}}{N_{2}},\cdots,\frac{K_{1n}}{N_{n}},N_{1}  \right)}\Z $, where $\gcd\left(a,b,\cdots\right)$ is the greatest common divisor of $a,b,\cdots$. In other words, for a deconfined particle excitation carrying $\Z_{N_1}$ gauge charges, the minimal nonzero amount of $\Z_{N_1}$ gauge charges is 
$		e_{1{\rm \min}}=\frac{N_{1}}{\gcd\left(\frac{K_{11}}{N_{1}},\frac{K_{12}}{N_{2}},\cdots,\frac{K_{1n}}{N_{n}},N_{1}\right)}.
 $ Since $e_1$ is equivalent to $e_1 +N_1$, the number of nonequivalent values of $e_1$ is $\gcd\left(\frac{K_{11}}{N_{1}},\frac{K_{12}}{N_{2}},\cdots,\frac{K_{1n}}{N_{n}},N_{1}\right)$, i.e., $e_1$ is labeled by $\Z_{\gcd\left(\frac{K_{11}}{N_{1}},\frac{K_{12}}{N_{2}},\cdots,\frac{K_{1n}}{N_{n}},N_{1}\right)}$.
This derivation can be applied to any $i$.  As a result, in order to make the particle labeled by $\mathbf{l}$ deconfined, all  $e_{i}$'s need to satisfy
\begin{equation}
	e_{i{\rm \min}}=\frac{N_{i}}{\gcd\left(\frac{K_{i1}}{N_{1}},\frac{K_{i2}}{N_{2}},\cdots,\frac{K_{in}}{N_{n}},N_{i}\right)}.
\end{equation}
Any particle excitation carrying $e_i$ units of $\Z_{N_i}$ gauge charges with $e_i \notin e_{i\min} \Z$ is confined. To illustrate, an example is shown in Fig. \ref{fig_higgs} where we consider a $BF$ theory with a single component $B B$ term with $\Z_{N_1}=\Z_{12}$ and $\frac{K_{11}}{N_1}=8$.
 
The confinement on $\Z_{N_i}$ gauge charge  also alters the period of $\Z_{N_i}$ gauge fluxes. Since the  $\Z_{N_i}$ gauge fluxes carried by a loop excitation can be detected by braiding a particle excitation around this loop excitation, we can consider the following particle-loop braiding phase: 
% {\textbf{\color{red}XXXXX check all symbols $\Theta,\Theta_{\mathrm{l}},\Theta_{PL},\Theta_{ex}, .....$}}{\color{blue}self-statistics: $\Theta_{l}$, mutual statistics: $\Theta_{l,l'}$, particle-loop braiding phase: $\Theta_{{\rm PL}}$, BR braiding phase: $\Theta_{\rm BR}$}
\begin{align}
	\Theta_{\rm{PL}}\left(e_{i\min},m_i\right)&=\exp \left[-\frac{\mathrm{i} 2\pi e_{i\min} m_{i}}{N_i}\right]\nonumber\\
	&=\exp \left[-\frac{\mathrm{i} 2\pi m_i}{\gcd\left(\frac{K_{11}}{N_{1}},\frac{K_{12}}{N_{2}},\cdots,\frac{K_{1n}}{N_{n}},N_{1}\right)}\right]. 
\end{align}
One can see that 
\begin{align}
	m_i \simeq m_i +\gcd\left(\frac{K_{11}}{N_{1}},\frac{K_{12}}{N_{2}},\cdots,\frac{K_{1n}}{N_{n}},N_{1}\right)
\end{align}
in the sense that $\Theta_{\rm{PL}}\left(e_{i\min},m_i\right)$ differs by an integral multiple of $2\pi$. An illustration for the smaller period of $m_i$ is presented in Fig. \ref{fig_higgs} where a $BF$ theory with a single component $B  B$ term with $\Z_{N_1}=\Z_{12}$ and $\frac{K_{11}}{N_1}=8$ is considered. In conclusion, the number of deconfined particle excitations and deconfined loop excitations are equivalent, satisfying the general belief of remote detectability of topological excitations in anomaly-free topological orders~\cite{Lan20180}.

\subsection{Self-statistics from the expectation values of  framed Wilson operators}
 Next, we study self-statistics (i.e., exchange statistics) of particle excitations in $3$D topological order which turns out to be controlled by the coefficients of the $BB$ term. Furthermore, the expression of self-statistics shares a similar form of that of $\left(2+1\right)$D Chern-Simons theory. 

In the following, we apply the standard methodology in TQFTs to determine self-statistics of particles: computing the expectation values of framed Wilson operators:
\begin{align}
	\!\!\langle W\left(\mathbf{l},\gamma\right)\rangle \!=\!\bigg\langle \!\exp\!\bigg(\int_{\gamma}{\rm i}\sum_{i=1}^{n}e_{i}A^{i}\!+\!\sum_{i,j=1}^{n}\frac{{\rm i}e_{i}K_{ij}}{N_{i}}\!\!\int_{\Sigma_{i}^{j}}B^{j}\!\bigg)\bigg\rangle\, , \nonumber
\end{align} 
where $\langle \mathcal{O}\rangle $ is defined as:
 $\langle \mathcal{O}\rangle = \mathcal{Z}^{-1}\int \mathcal{D}A^i\mathcal{D}B^i \exp({\rm i}S)\mathcal{O}\, $.  The partition function $\mathcal{Z}=\int  \mathcal{D}A^i\mathcal{D}B^i \exp({\rm i}S)$ with the action given by Eq.~(\ref{eq:n_component_BB term}).

\begin{figure}
	\centering
	\includegraphics[width=0.4\textwidth]{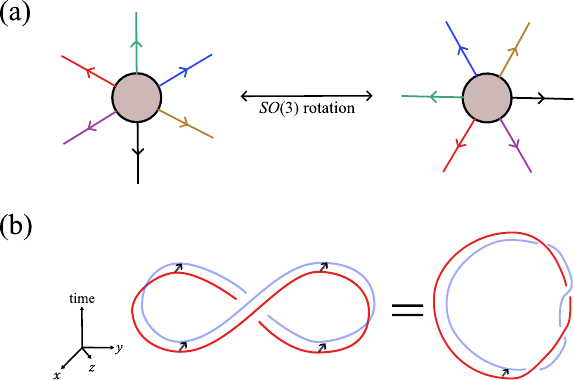}
	\caption{(a) The physical picture of $B   B$ term is to bind flux strings (lines with arrow) to particle excitations (red solid circle), see the Wilson operator of a particle excitation (\ref{eq:op_particle_n_component}). %%The bound state of particle excitation and the endpoint %%of the attached flux strings is no longer isotropic in %%$3$D space. 
	Different orientations of flux strings can be connected by an $SO\left(3\right)$ rotation.  (b) Consider a particle excitation attached by a flux string, e.g., Eq. (\ref{eq:op_particle_single_component}), exchanging such two particle excitations can be viewed as a self $2\pi$ rotation of a particle. A nontrivial framing of the world line of particle excitation is shown to illustrate this point.}
	\label{fig_framing}
\end{figure}

For this purpose, we integrate out $A^{i}$, which   
results in
 $	B^{i}=-\frac{2\pi e_{i}}{N_{i}}\delta^{\perp}\left(\Sigma_{i}\right)
 $ 
with $\partial\Sigma_{i}=\gamma$. $\delta^{\perp}\left(\Sigma_{i}\right)$ is a delta distribution supported on $\Sigma_i$, which is $2$-form valued since $B^i$ is a $2$-form. Plugging this solution back to the
path integral, we have 
\begin{align}
	\left\langle W\left(\mathbf{l},\gamma\right)\right\rangle = & \exp\left[{\rm i}\sum_{i,j=1}^{n}\frac{K_{ij}}{4\pi}\frac{2\pi e_{i}}{N_{i}}\frac{2\pi e_{j}}{N_{j}}\#\left(\Sigma_{i}\cap\Sigma_{j}\right)\right]\nonumber \\
	& \times\exp\left[-{\rm i}\sum_{i,j=1}^{n}\frac{{\rm i}e_{i}K_{ij}}{N_{i}}\frac{2\pi e_{j}}{N_{j}}\cdot\#\left(\Sigma_{i}^{j}\cap\Sigma_{j}\right)\right]
\end{align}
where $\#\left(\Sigma_{i}^{j}\cap\Sigma_{j}\right)$ is the intersection number of two Seifert surfaces $\Sigma_{i}^{j}$ and $\Sigma_{j}$. It equals to $1$ if $\gamma$ has a nontrivial framing, see Fig.~\ref{fig_framing}(b). Using $\#\left(\left(\Sigma_{i}-\Sigma_{i}^{j}\right)\cap\left(\Sigma_{j}-\Sigma_{j}^{i}\right)\right)=0$, which is because two closed manifolds (i.e., the difference of two Seifert surfaces) in $S^4$ has zero intersection number, one has 
\begin{align}
&\#\left(\Sigma_{i}\cap\Sigma_{j}\right)-\#\left(\Sigma_{i}^{j}\cap\Sigma_{j}\right)-\#\left(\Sigma_{i}\cap\Sigma_{j}^{i}\right)\nonumber\\
	=&-\#\left(\Sigma_{i}^{j}\cap\Sigma_{j}^{i}\right).
\end{align}

Therefore we have 
\begin{equation}
	\left\langle W\left(\mathbf{l},\gamma\right)\right\rangle =\exp\left[-{\rm i}\sum_{i,j=1}^{n}\frac{\pi K_{ij}e_{i}e_{j}}{N_{i}N_{j}}\#\left(\Sigma_{i}^{j}\cap\Sigma_{j}^{i}\right)\right].
\end{equation}
$\#\left(\Sigma_{i}^{j}\cap\Sigma_{i}^{j}\right)=1$ if a nontrivial framing is
introduced. The self-statistics of a particle with $e_{i}$ charge
is $\exp\left(-{\rm i}\frac{\pi K_{ii}e_{i}e_{i}}{N_{i}N_{i}}\right)$.
Off-diagonal terms $\exp\left(-{\rm i}\frac{2\pi K_{ij}e_{i}e_{j}}{N_{i}N_{j}}\right)$
with $i\neq j$ is the mutual statistics of two particle excitations
with $e_{i}$ units of $\mathbb{Z}_{N_{i}}$ gauge charges and $e_{j}$
units of $\mathbb{Z}_{N_{j}}$ gauge charges. Remember that for a deconfined particle excitation, $e_{i}\in e_{i\min}\mathbb{Z}\mod N_{i}$, such $e_{i}$'s guarantee that the self-statistics of a particle is $\pm1$ and the mutual statistics of two particles are always trivial. In conclusion, for a
particle labeled by $\mathbf{l}=\left(e_{1},e_{2},\cdots,e_{n}\right)^{T}$,
the self (exchange) statistics is given by
\begin{equation}
	\Theta_{\mathbf{l}}=\exp\left(-{\rm i}\sum_{i,j=1}^{n}\frac{\pi K_{ij}e_{i}e_{j}}{N_{i}N_{j}}\right)=\exp\left({\rm -i}\pi\mathbf{l}^{T}\widetilde{K}\mathbf{l}\right)
	\label{eq_general_formula_self}\,,
\end{equation}
where $\left(\widetilde{K}\right)_{ij}=\frac{K_{ij}}{N_{i}N_{j}}$.
One can recognize that this result is similar to the self-statistics of  particles in $\left(2+1\right)$D Chern-Simons theory. In the Chern-Simons theory with a $K_{\rm CS}$ matrix, the self-statistics of a particle labeled by a vector $\mathbf{l}^T$ is characterized by  
\begin{align}
	\Theta_{\mathbf{l}}^{\rm CS}=\exp\left({\rm i}\pi \mathbf{l}^T K_{\rm CS} \mathbf{l}\right).
\end{align}
In addition, when all diagonal elements of $K$ are \emph{even}, the trivial particle excitation [$\mathbf{l}=\left(0 \mod N_1, \cdots, 0 \mod N_n\right)^T$] is bosonic. When at least one diagonal element is \emph{odd}, this theory admits fermionic trivial particle excitation. This result is similar to that in the $K_{\rm CS}$ Chern-Simons theory.  
The coefficient matrix $K$ of the $B B$ term plays a similar role as that of the Chern-Simons theory.

So far, we have seen how a $K$-matrix $B  B$ term
dramatically changes the number of deconfined operators and exchange
statistics of a $\prod_{i=1}^{n}\mathbb{Z}_{N_{i}}$ gauge theory. The exchange and mutual statistics can be better explained by the
examples of the $BF$ theory with a single (two-) component $B  B$
term. In the single component case, the action is 
\begin{equation}
	S=\int\frac{N_{1}}{2\pi}B^{1}dA^{1}+\frac{K_{11}}{4\pi}B^{1}B^{1}
	\label{eq_action_BF+single_BB}
\end{equation}
and the Wilson operator of a particle excitation carrying $e_{1}$
units of $\mathbb{Z}_{N_{1}}$ gauge charges is 
\begin{align}
	W\left(e_{1},\gamma\right)= & \exp\left({\rm i}e_{1}\int_{\gamma_{1}}A^{1}+\frac{{\rm i}e_{1}K_{11}}{N_{1}}\int_{\Sigma_{1}}B^{1}\right),
	\label{eq:op_particle_single_component}
\end{align}
which describe a particle excitation with one attached flux string.
For a deconfined particle excitation, it is required that 
\begin{equation}
	e_{1\min}=\frac{N_{1}}{\gcd\left(\frac{K_{11}}{N_{1}},N_{1}\right)}.
\end{equation}
To calculate self-statistics of this particle excitation, we can
make use of spin-statistics theorem, see Fig.~\ref{fig_framing}(b). Its expectation value is
\begin{equation}
	\left\langle W\left(e_{1},\gamma\right)\right\rangle =\exp\left[-\frac{{\rm i}\pi K_{11}e_{1}e_{1}}{N_{1}N_{1}}\cdot\#\left(\Sigma_{1}\cap\Sigma_{1}\right)\right].
\end{equation}
The value of $\#\left(\Sigma_{1}\cap\Sigma_{1}\right)$ depends on
whether the framing of $\gamma$ is nontrivial or not. A framing of
$\gamma$ can be understood as assigning a vector on each point along
$\gamma$. Actually, we are now considering a particle attached with
a flux string. In regularization, the charge and the endpoint of flux string (i.e., monopole) cannot be
placed on the same lattice site, i.e., the charge-monopole composite is
not isotropic. It is necessary to use a vector to indicate the shape
of the  composite. Such vectors along the world line of
particle constitute the framing. In $\left(2+1\right)$D, there
are different ways to equip a vector to each point along the world
line. The number of ways to equip is $\pi_{1}\left(SO\left(2\right)\right)=\mathbb{Z}$
that counts nonequivalent mappings from $S_{1}$ (the world line)
to $SO\left(2\right)$ (2D rotation of vector on each point). $\pi_{1}\left(SO\left(2\right)\right)=\mathbb{Z}$
means that in $\left(2+1\right)$D there can be anyonic statistics.
In $\left(3+1\right)$D, the $3$D rotation of vector on each point
is captured by $SO\left(3\right)$ and $\pi_{1}\left(SO\left(3\right)\right)=\mathbb{Z}_{2}$
means that there are only two kinds of statistics in $\left(3+1\right)$D.

\begin{table*}[t]
	\caption{Properties of $S=\int\frac{N_{1}}{2\pi}B^{1}  dA^{1}+\frac{K_{11}}{4\pi}B^{1}  B^{1}$
		with different values of $N_{1}$ and $K_{11}$. $\Theta_{\text{trivial}}$
		 is the self-statistics of trivial particle excitation, i.e., those carrying $0\mod N$ unit of gauge charge. $\Theta_{\text{trivial}}=1\left(-1\right)$ means that the trivial particle excitation is a boson (fermion), or equivalently, the theory is a bosonic (fermionic) one. $\Theta_{e_{1\min}}$ is the self-statistics of a particle excitation with $e_{1\min}$ units of gauge charge where $e_1\min=N_1 / \gcd\left(\frac{K_11}{N_1},N_1\right)$. $e_{1\min}$ is the minimal units of gauge charge carried by a deconfined particle excitation in theory $S=\int\frac{N_{1}}{2\pi}B^{1}  dA^{1}+\frac{K_{11}}{4\pi}B^{1}  B^{1}$. An emergent fermion appears when $\Theta_\text{trivial}=1$ while $\Theta_{e_{1\min}}=-1$. In other words, there exists nontrivial fermionic particle excitations in a bosonic theory. \label{table_singleBB_properties}}
	\begin{tabular*}{\textwidth}{@{\extracolsep{\fill}}cccccc}
		\toprule 
		\hline
		
		\hline
		
		$N_{1}$ & $K_{11}$ & $\Theta_{\text{trivial}}$ & $\Theta_{e_{1\min}}$ & bosonic/fermionic theory? & emergent fermion? \\\hline
		\midrule
		odd & even & $1$ & $1$ & bosonic & No\tabularnewline
		odd & odd & $-1$ & $-1$ & fermionic & -\tabularnewline
		even & odd & $1$ & $1$ & bosonic & No\tabularnewline
		even & even and $\frac{{\rm lcm}\left(\frac{K_{11}}{N_{1}},N_{1}\right)}{\gcd\left(\frac{K_{11}}{N_{1}},N_{1}\right)}\in2\mathbb{Z}$ & $1$ & $1$ & bosonic & No\tabularnewline
		even & even and $\frac{{\rm lcm}\left(\frac{K_{11}}{N_{1}},N_{1}\right)}{\gcd\left(\frac{K_{11}}{N_{1}},N_{1}\right)}\in2\mathbb{Z}+1$ & $1$ & $-1$ & bosonic & Yes\tabularnewline
		\bottomrule
		\hline
		
		\hline
		
	\end{tabular*}
\end{table*}

For a nontrivial framing of $\gamma$, $\#\left(\Sigma_{1}\cap\Sigma_{1}\right)=1$,
which also indicates a $2\pi$-rotation of this particle excitation
that induces a phase 
\begin{equation}
	\Theta_{e_1}=\exp\left(-\frac{{\rm i}\pi K_{11}e_{1}e_{1}}{N_{1}N_{1}}\right).
\end{equation}
According to spin-statistics theorem, $\Theta_{e_1}$
is the self-statistics of particle excitation with $e_{1}$ units
of gauge charge. Notice that $e_{1}\in\frac{N_{1}}{\gcd\left(\frac{K_{11}}{N_{1}},N_{1}\right)}\mathbb{Z}$,
we find $\Theta_{e_{1\min}}=\exp\left[-\frac{{\rm i}\pi{\rm lcm}\left(\frac{K_{11}}{N_{1}},N_{1}\right)}{\gcd\left(\frac{K_{11}}{N_{1}},N_{1}\right)}\right]=\pm1$
corresponding to bosonic or fermionic statistics.

For a trivial particle excitation, i.e., that with $e_{1}=0\mod N_{1}$,
its self-statistics is given by 
\begin{equation}
	\Theta_{{\rm trivial}}=\exp\left(-{\rm i}\pi K_{11}\right).
\end{equation}
When $K_{11}$is odd, $\Theta_{{\rm trivial}}=-1$ meaning the trivial
particle excitation is fermionic which tells us the theory (\ref{eq_action_BF+single_BB})
is a fermionic theory. Notice that an odd $K_{11}$ can only happen
when the theory is defined on a spin manifold. When $K_{11}$ is even,
$\Theta_{\rm trivial}=1$ indicating that the trivial particle excitation is a boson,
i.e., the theory (\ref{eq_action_BF+single_BB}) is a bosonic
one.

For nontrivial particle excitations, i.e., those with $e_{1}\neq0\mod N_{1}$,
their self-statistics depends on the values of $N_{1}$, $\frac{K_{11}}{N_{1}}$,
and $n$. Among all possible combinations, it is possible that some
particle excitations with $e_{1}\neq0\mod N_{1}$ are fermionic while
the trivial one is bosonic. We call such particle excitations \emph{emergent
	fermions} in the sense that they exhibit fermionic statistics in a
bosonic theory. Below we summary the properties of theory (\ref{eq_action_BF+single_BB})
for different $N_{1}$ and $K_{11}$ in Table \ref{table_singleBB_properties}.

The second example is a two-component $B  B$ term with the action
is 
\begin{equation}
	S=\int\sum_{i=1}^{2}\frac{N_{i}}{2\pi}B^{i}dA^{i}+\sum_{i,j=1}^{2}\frac{K_{ij}}{4\pi}B^{i}B^{j}.\label{eq:2-component_BB_term-1}
\end{equation}
Consider a particle excitation carrying two types of gauge charges,
denoted by $\mathbf{l}=\left(e_{1},e_{2}\right)^{T}$, its exchange
statistics is given by 
\begin{align}
	\left\langle W\left(\mathbf{l},\gamma\right)\right\rangle = & \frac{1}{\mathcal{Z}}\int{D}\left[A^{i}, B^{i}\right]\exp\left({\rm i}S\right)\nonumber \\
	& \times\exp\left(\int_{\gamma}{\rm i}\sum_{i=1}^{2}e_{i}A^{i}+\sum_{i=1}^{2}\sum_{j=1}^{2}\frac{{\rm i}e_{i}K_{ij}}{N_{i}}\int_{\Sigma_{i}^{j}}B^{j}\right)\nonumber \\
	= & \exp\left[-{\rm i}\sum_{i,j=1}^{2}\frac{\pi K_{ij}e_{i}e_{j}}{N_{i}N_{j}}\#\left(\Sigma_{i}^{j}\cap\Sigma_{j}^{i}\right)\right].
\end{align}
By choosing a nontrivial framing of $\gamma$, i.e., $\#\left(\Sigma_{i}^{j}\cap\Sigma_{j}^{i}\right)=1$,
we find the self-statistics of a particle excitation labeled by
$\mathbf{l}=\left(e_{1},e_{2}\right)^{T}$ is 
\begin{equation}
	\Theta_{\mathbf{l}}=\exp\left(-\frac{{\rm i}\pi K_{11}e_{1}e_{1}}{N_{1}N_{1}}-\frac{{\rm i}\pi K_{22}e_{2}e_{2}}{N_{2}N_{2}}-\frac{{\rm i}2\pi K_{12}e_{1}e_{2}}{N_{1}N_{2}}\right),
\end{equation}
where $\exp\left(-\frac{{\rm i}2\pi K_{12}e_{1}e_{2}}{N_{1}N_{2}}\right)$
is the mutual statistics of $\mathbb{Z}_{N_{1}}$ charges and $\mathbb{Z}_{N_{2}}$
charges. Keep in mind that for a deconfined particle excitation, $e_{i}\in q_{i\min}\mathbb{Z}\mod N_{i}$
where $e_{i\min}=\frac{N_{i}}{\gcd\left(\frac{K_{i1}}{N_{1}},\frac{K_{i2}}{N_{2}},N_{i}\right)}$.
Such $e_{i}$'s guarantee that the self-statistics of a particle
is $\pm1$ and the mutual statistics of two particles are always trivial.
To see this, we consider 
\begin{align}
	\Theta_{e_{1\min},e_{2\min}} & =\exp\left(-\frac{{\rm i}2\pi K_{12}e_{1\min}e_{2\min}}{N_{1}N_{2}}\right).
\end{align}
We can see that 
\begin{align}
	\frac{K_{12}e_{1\min}e_{2\min}}{N_{1}N_{2}}= & \frac{K_{12}}{N_{1}N_{2}}\frac{N_{1}}{\gcd\left(\frac{K_{11}}{N_{1}},\frac{K_{12}}{N_{2}},N_{1}\right)}\frac{N_{2}}{\gcd\left(\frac{K_{21}}{N_{1}},\frac{K_{22}}{N_{2}},N_{2}\right)}\nonumber \\
	\in & \frac{K_{12}}{N_{1}N_{2}}\frac{N_{1}}{\gcd\left(\frac{K_{12}}{N_{2}},N_{1}\right)}\frac{N_{2}}{\gcd\left(\frac{K_{21}}{N_{1}},N_{2}\right)}\mathbb{Z}\\
	\nonumber 
\end{align}
since $\gcd\left(\frac{K_{12}}{N_{2}},N_{1}\right)\in\gcd\left(\frac{K_{11}}{N_{1}},\frac{K_{12}}{N_{2}},N_{1}\right)\mathbb{Z}$
and $\gcd\left(\frac{K_{21}}{N_{1}},N_{2}\right)\in\gcd\left(\frac{K_{21}}{N_{1}},\frac{K_{22}}{N_{2}},N_{2}\right)\mathbb{Z}$.
Furthermore, using
\begin{align}
	& \frac{K_{12}}{N_{1}N_{2}}\frac{N_{1}}{\gcd\left(\frac{K_{12}}{N_{2}},N_{1}\right)}\frac{N_{2}}{\gcd\left(\frac{K_{21}}{N_{1}},N_{2}\right)}\nonumber \\
	= & \frac{K_{12}N_{1}N_{2}}{\gcd\left(K_{12},N_{1}N_{2}\right)\gcd\left(K_{21},N_{1}N_{2}\right)}\nonumber \\
	= & \frac{{\rm lcm}\left(K_{12},N_{1}N_{2}\right)}{\gcd\left(K_{12},N_{1}N_{2}\right)}
\end{align}
we can see that $\Theta_{{\rm mutual}}\left(e_{1\min},e_{2\min}\right)=\exp\left(-{\rm i}2\pi\mathbb{Z}\right)=1$.
This is consistent with the fact that in $3$D space the mutual statistics (i.e., full braiding)
of two particles is topologically trivial.

\section{TQFT with nontrivial braiding statistics and emergent fermions}\label{section_compatible}
  
\subsection{Particle-loop braiding in the presence of emergent fermions ($BF+BB$)}\label{sec_plbraid}

A pure $BF$ theory describes the particle-loop braiding. A $BF$
term is compatible with a $BB$ term to form a legitimate TQFT
action. This simplest $BF$ theory with a single component $BB$ term is given by Eq.~(\ref{eq_action_BF+single_BB}). Emergent fermionic particle excitations are possible provided
proper values of $N_{1}$ and $K_{11}$. To explicitly show how emergent
fermion influences particle-loop braiding, we can consider the phase
of particle-loop braiding given by 
\begin{align}
	\Theta_{\rm PL}= & \frac{1}{\mathcal{Z}}\int\mathcal{D}\left[A^{i}\right]\mathcal{D}\left[B^{i}\right]\exp\left({\rm i}S\right)\nonumber \\
	& \times\exp\left({\rm i}e_{1}\int_{\gamma}A^{1}+{\rm i}\frac{e_{1}K_{11}}{N_{1}}\int_{\Sigma}B^{1}\right)\exp\left({\rm i}m_{1}\int_{\sigma}B^{3}\right)
\end{align}
where $\gamma$ is a closed curve with $\partial\Sigma=\gamma$, $\sigma$
is a closed surface, $e_{1}$ and $m_{1}$ are the numbers of charges
and fluxes carried by the particle and the loop. $\gamma$ and $\sigma$
can be understood as the world line and world sheet of the particle
and the loop. The phase of particle-loop braiding is 
\begin{equation}
	\Theta_{\rm PL}=\exp\left[-\frac{{\rm i}\pi K_{11}e_{1}e_{1}}{N_{1}N_{1}}\#\left(\Sigma\cap\Sigma\right)-\frac{{\rm i}2\pi e_{1}m_{1}}{N_{1}}\#\left(\Sigma'\cap\sigma\right)\right]
\end{equation}
where $\partial\Sigma'=\gamma$ and $\#\left(\Sigma'\cap\sigma\right)$
is the linking number of $\gamma$ and $\sigma$. There are two contributions
to this phase. $\exp\left[-\frac{{\rm i}2\pi e_{1}m_{1}}{N_{1}}\#\left(\Sigma'\cap\sigma\right)\right]$
is the usual particle-loop braiding phase due to the particle traveling
around the loop. $\exp\left[-\frac{{\rm i}\pi K_{11}e_{1}e_{1}}{N_{1}N_{1}}\#\left(\Sigma\cap\Sigma\right)\right]$
is just the self-statistics of the particle excitation.  As discussed in previous
section, the values of $e_{1}$ and $m_{1}$ are constrained by
\begin{align}
	e_{1}= & e_{1\min}\cdot p=\frac{N_{1}}{\gcd\left(\frac{K_{11}}{N_{1}},N_{1}\right)}\cdot p,p\in\mathbb{Z}_{\gcd\left(\frac{K_{11}}{N_{1}},N_{1}\right)};\\
	m_{1}\simeq & m_{1}+\gcd\left(\frac{K_{11}}{N_{1}},N_{1}\right).
\end{align}
Consider a particle and a loop carrying minimal gauge charge and flux,
the phase contributed by a particle-loop braiding is given by 
\begin{equation}
	\Theta_{{\rm PL}}=\exp\left(-\frac{{\rm i}2\pi e_{1\textrm{min}}m_{1}}{N_{1}}\right)=\exp\left(-\frac{{\rm i}2\pi m_{1}}{\gcd\left(\frac{K_{11}}{N_{1}},N_{1}\right)}\right)
\end{equation}
where $m_{1}\in\mathbb{Z}_{\gcd\left(\frac{K_{11}}{N_{1}},N_{1}\right)}$.
This means that the $BF$ theory with a nontrivial $BB$ term only labels
fewer topologically ordered phases than a pure $BF$ theory. This
is because a $BB$ term would confine part of topological excitations,
making the physical observable braiding phases fewer.

 Each topological
excitation $\mathsf{e}$ can be represented by a gauge invariant Wilson
operator $\mathcal{O}_{\mathsf{e}}$. Using path integral, we can
extract fusion rules $\mathsf{a}\otimes\mathsf{b}=\oplus_{i}N_{\mathsf{e}_{i}}^{\mathsf{a}\mathsf{b}}\mathsf{e}_{i}$
from \citep{PhysRevB.107.165117}
\begin{align}
	\left\langle \mathsf{a}\otimes\mathsf{b}\right\rangle = & \frac{1}{\mathcal{Z}}\int\mathcal{D}\left[A^{i},B^{i}\right]\exp\left({\rm i}S\right)\times\left(\mathcal{O}_{\mathsf{a}}\times\mathcal{O}_{\mathsf{b}}\right)\nonumber \\
	= & \frac{1}{\mathcal{Z}}\int\mathcal{D}\left[A^{i},B^{i}\right]\exp\left({\rm i}S\right)\times\left(\sum_{i}N_{\mathsf{e}_{i}}^{\mathsf{a}\mathsf{b}}\mathcal{O}_{\mathsf{e}_{i}}\right)\nonumber \\
	= & \left\langle \oplus_{i}N_{\mathsf{e}_{i}}^{\mathsf{a}\mathsf{b}}\mathsf{e}_{i}\right\rangle .
	\label{eq:fusion_path_integral}
\end{align}
Since emergent fermion can be induced by a proper $BB$ term,
we can couple a $BB$ term to other topological terms such
that we can study whether and how emergent fermion would influence
braiding statistics and fusion rules.

Consider a general topological excitation labeled by $\left(e_{1},m_{1}\right)$,
when $m_{1}=0$ it is a point like particle excitation; when $e_{1}=0$,
it is a pure loop excitation (a loop excitation without particle attached
	on it); when $e_{1},m_{1}\neq0$, it is a decorated loop excitation, i.e.,
the bound state of a particle and a pure loop. It is straightforward
to see that the fusion rule of two topological excitations is given
by 
\begin{equation}
	\left(e_{1},m_{1}\right)\otimes\left(e_{1}',m_{1}'\right)=\left(e_{1}+e_{1}',m_{1}+m_{1}'\right).
\end{equation}
Since both $e_{1}$ and $m_{1}$ are labeled by $\mathbb{Z}_{\gcd\left(\frac{K_{11}}{N_{1}},N_{1}\right)}$,
the fusion rules can be captured by a $\mathbb{Z}_{\gcd\left(\frac{K_{11}}{N_{1}},N_{1}\right)}\times\mathbb{Z}_{\gcd\left(\frac{K_{11}}{N_{1}},N_{1}\right)}$
group. While for a pure $BF$ theory $S=\int\frac{N_{1}}{2\pi}B^{1}dA^{1}$,
its fusion rules are captured by a $\mathbb{Z}_{N_{1}}\times\mathbb{Z}_{N_{1}}$
group. 

In conclusion, the coefficient of $BB$ term $K_{11}$ would
confine, either partially or completely, the gauge group structure of $\mathbb{Z}_{N_{1}}$$BF$ theory,
leaving the deconfined gauge group to be $\mathbb{Z}_{\gcd\left(\frac{K_{11}}{N_{1}},N_{1}\right)}$.
Only $\gcd\left(\frac{K_{11}}{N_{1}},N_{1}\right)$of $N_{1}$ particle
excitations are deconfined and the gauge fluxes are labeled by $\mathbb{Z}_{\gcd\left(\frac{K_{11}}{N_{1}},N_{1}\right)}$.
The cyclic structure of particle-loop braiding phase is described
by the deconfined gauge group, so are the fusion rules. The fusion rules
are still Abelian. 

\subsection{Multi-loop braiding   in the presence of emergent fermions ($BF+AAdA/AAAA+BB$)}\label{sec_mlbraid}

In $3$D topological order, multi-loop braiding includes three-loop
braiding (described by an $AAdA$ topological term) and four-loop
braiding (described by an $AAAA$ topological term). The corresponding
simplest TQFT actions are as follows: for a three-loop braiding: 
\begin{equation}
	S_{{\rm 3L}}=\int\sum_{i=1}^{2}\frac{N_{i}}{2\pi}B^{i}dA^{i}+q_{1}A^{1}A^{2}dA^{2}
\end{equation}
or 
\begin{equation}
	S_{{\rm 3L}}'=\int\sum_{i=1}^{2}\frac{N_{i}}{2\pi}B^{i}dA^{i}+q_{2}A^{2}A^{1}dA^{1}
\end{equation}
with $G=\mathbb{Z}_{N_{1}}\times\mathbb{Z}_{N_{2}}$ and a proper
coefficient $q_{1},q_{2}$; for a four-loop braiding: 
\begin{equation}
	S_{{\rm 4L}}=\int\sum_{i=1}^{4}\frac{N_{i}}{2\pi}B^{i}dA^{i}+q_{\textrm{4L}}A^{1}A^{2}A^{3}A^{4}
\end{equation}
with $G=\prod_{i=1}^{4}\mathbb{Z}_{N_{i}}$ and a proper coefficient
$q_{{\rm 4L}}$. Now we try to consider emergent fermion together
with multi-loop braiding. Based on discussion in previous section,
we want to add a $BB$ term to above topological actions for three-loop
braiding or four-loop braiding. However, according to the condensation
picture illustrated in Sec. \ref{subsec_condensation_picture}, such
attempt would not success. The topological action for a multi-loop
braiding originates from a multi-layer Abel-Higgs model in which each
layer describes a condensate of \emph{boson} coupled to $1$-form
gauge field ($A^{i}$). For example, an $A^{1}A^{2}dA^{2}$ topological
term is derived from the interaction of two layers of boson condensation.
Introducing a $BB$ term requires that at least one layer of is vortex-line
condensation. Since a boson condensation is totally different from
a vortex-line condensation, it is impossible for a Abel-Higgs model
to describe both of them simultaneously. In other words, one of $A^{i}$
and $B^{i}$ must be Lagrange multiplier and they cannot both appear
in   $AAdA+BB$. We may draw such a conclusion: in
bosonic topological orders, multi-loop braiding is not compatible
with emergent fermion, based on our condensation picture. 
%On the other
 
In principle, there
seems no reason for forbiding multi-loop braidings in a system
that supports emergent fermions. For example, multi-loop braiding is studied
in gauged fermionic symmetry-protected topological (fSPT) phase, see, e.g.,
Refs. \citep{loop_statistics_fSPT,2021NatnonAbelian3loop}. Some lattice
cocycle models are used to describe multi-loop braiding in fermionic
system \citep{loop_statistics_fSPT}.  A question arises:
how to describe the coexistence of multiloop braidings and emergent fermions in continuum field theory that is believed
to be capable for liquid-like phases of matter that have   well-defined thermodynamical limit? We leave this question to the future exploration. Here, we come up with some hints: while Wilson operators for fermions should be regularized by framing, loop excitations that correspond to emergent fermions in a given gauge subgroup $\Z_{N_i}$ may also require a regularization of some sort.

\subsection{Borromean rings braiding in the presence of emergent fermions ($BF+AAB+BB$)}\label{sec_brbraid}

A Borromean rings braiding is described by an $AAB$ topological term
\citep{yp18prl}. A system is equipped with Borromean rings topological
order if it supports a Borromean rings braiding. A Borromean rings
topological order is featured by non-Abelian fusion rules and loop
shrinking rules \citep{PhysRevB.107.165117}. Unlike $AAdA$ term, there is a chance that an
$AAB$ term can be compatible with $BB$ term, so we can consider
the following TQFT action:
\begin{equation}
	S=\int\sum_{i=1}^{3}\frac{N_{i}}{2\pi}B^{i}dA^{i}+qA^{1}A^{2}B^{3}+\frac{K_{33}}{4\pi}B^{3}B^{3}\label{eq:action_AAB+BB}\,,
\end{equation}
where $q=\frac{pN_{1}N_{2}N_{3}}{N_{123}}$ with $N_{123}=\gcd\left(N_{1},N_{2},N_{3}\right)$,
$p\in\mathbb{Z}_{N_{123}}$, and the gauge group is $G=\prod_{i=1}^{3}\mathbb{Z}_{N_{i}}$.
The gauge transformations are 
\begin{align}
	A^{1}\rightarrow & A^{1}+d\chi^{1},\\
	A^{2}\rightarrow & A^{2}+d\chi^{2},\\
	A^{3}\rightarrow & A^{3}+d\chi^{3}-\frac{K_{33}}{N_{3}}V^{3}\nonumber \\
	& -\frac{2\pi q}{N_{3}}\left(\chi^{1}A^{2}+\frac{1}{2}\chi^{1}d\chi^{2}\right)\nonumber \\
	& +\frac{2\pi q}{N_{3}}\left(\chi^{2}A^{1}+\frac{1}{2}\chi^{2}d\chi^{1}\right),\\
	B^{1}\rightarrow & B^{1}+dV^{1}\nonumber \\
	& -\frac{2\pi q}{N_{1}}\left(\chi^{2}B^{3}-A^{2}V^{3}+\chi^{2}dV^{3}\right),\\
	B^{2}\rightarrow & B^{2}+dV^{2}\nonumber \\
	& +\frac{2\pi q}{N_{2}}\left(\chi^{1}B^{3}-A^{1}V^{3}+\chi^{1}dV^{3}\right),\\
	B^{3}\rightarrow & B^{3}+dV^{3}.
\end{align}
The compatibility of $AAB$ term and $BB$ term indicates that emergent
fermion is possible in Borromean rings topological order.

We use $\mathsf{P}_{e_{1}e_{2}e_{3}}$ to denote a particle excitation
carrying $e_{i}$ units of $\mathbb{Z}_{N_{i}}$ gauge charges and
$\mathsf{L}_{m_{1}m_{2}m_{3}}$to denote a pure loop excitation carrying
$m_{i}$ units of $\mathbb{Z}_{N_{i}}$ gauge fluxes ($i=1,2,3$).
A decorated loop (formed by attaching a particle excitation to a pure
loop excitation) is denoted by $\mathsf{L}_{m_{1}m_{2}m_{3}}^{e_{1}e_{2}e_{3}}$.
Consider a Borromean rings braiding involving $\mathsf{L}_{m_{1}00}$,
$\mathsf{L}_{0m_{2}0}$, and $\mathsf{P}_{00e_{3}}$, the phase is
\begin{align}
	\Theta_{{\rm BR}}\left(m_{1},m_{2},e_{3}\right)= & \exp\left[-\frac{{\rm i}2\pi pm_{1}m_{2}e_{3}}{N_{123}}\cdot{\rm Tlk}\right]\nonumber \\
	& \times\exp\left[-\frac{{\rm i}\pi K_{33}e_{3}e_{3}}{N_{3}N_{3}}\#\left(\Sigma\cap\Sigma\right)\right]
\end{align}
where ${\rm Tlk}$ is the Milnor's triple linking number of the link
formed by the two loops' world sheets $\sigma_{1},\sigma_{2}$ and
the particle's world line $\gamma$, $\Sigma$ is a Seifert surface
of $\gamma$. The first term is the phase of Borromean rings braiding
$\Theta_{{\rm BR}}$ \citep{yp18prl} while the second term is due
to the possible self $2\pi$ rotation of $\mathsf{P}_{00e_{3}}$ during
the braiding process. Since the self $2\pi$ rotation of $\mathsf{P}_{00e_{3}}$
would introduce an extra phase of $\pm1$, depending on its own exchange
statistics (spin-statistics theorem), we can just ignore it. Notice
that the existence of $BB$ term will confine some particle
excitation carrying $\mathbb{Z}_{N_{3}}$gauge charges, the value
of $e_{3}$ is given by 
\begin{equation}
	e_{3}\in\frac{N_{3}}{\gcd\left(\frac{K_{33}}{N_{3}},N_{3}\right)}\mathbb{Z}.
\end{equation}
When $K_{33}=0$, i.e., no $BB$ term considered, $e_{3}$
is labeled by $\mathbb{Z}_{N_{3}}$ and the minimal $e_{3}$ is $1$.
The Borromean rings braiding phase 
\begin{equation}
	\Theta_{{\rm BR}}\left(m_{1},m_{2},1\right)=\exp\left(-\frac{{\rm i}2\pi pm_{1}m_{2}}{N_{123}}\right)
\end{equation}
is labeled by $p\in\mathbb{Z}_{N_{123}}$. In the case of $K_{33}\neq0$,
the minimal $e_{3}$ cannot be $1$ any longer since it would be confined.
The minimal value of $e_{3}$ is $e_{3\min}=\frac{N_{3}}{\gcd\left(\frac{K_{33}}{N_{3}},N_{3}\right)}$.
The Borromean rings braiding phase is 
\begin{align}
	\Theta_{{\rm BR}}\left(m_{1},m_{2},e_{3\min}\right) & =\exp\left(-\frac{{\rm i}2\pi pm_{1}m_{2}e_{3\min}}{N_{123}}\right)\nonumber \\
	& =\exp\left(-\frac{{\rm i}2\pi pm_{1}m_{2}N_{3}}{N_{123}\gcd\left(\frac{K_{33}}{N_{3}},N_{3}\right)}\right).
\end{align}
Since $\frac{N_{3}}{N_{123}}\in\mathbb{Z}$, we can see that $p$
is identified with $p+\gcd\left(\frac{K_{33}}{N_{3}},N_{3}\right)$.
Combined with $p\in\mathbb{Z}_{N_{123}}$, we find that $p$ is actually
labeled by $\mathbb{Z}_{\gcd\left(N_{1},N_{2},N_{3},\frac{K_{33}}{N_{3}}\right)}$.
We can see that adding a $BB$ term to $S=\int\sum_{i=1}^{3}\frac{N_{i}}{2\pi}B^{i}dA^{i}+qA^{1}A^{2}B^{3}$
may reduce the number of different Borromean rings braiding phases.
This result is reasonable since some particle excitations are confined
by $BB$ term hence cannot contribute to an observable Borromean
rings braiding phases .

First, let us find out Wilson operators for those topological excitation
carrying only \emph{one} kind of gauge charge or flux for the action
\begin{equation}
	S=\int\sum_{i=1}^{3}\frac{N_{i}}{2\pi}B^{i}dA^{i}+qA^{1}A^{2}B^{3}+\frac{K_{33}}{4\pi}B^{3}B^{3}\label{eq:action_AAB+BB_z2z2z6-1}
\end{equation}
with $G=\mathbb{Z}_{N_{1}}\times\mathbb{Z}_{N_{2}}\times\mathbb{Z}_{N_{3}}$.
The particle excitations carrying $\mathbb{Z}_{N_{1}}$ or $\mathbb{Z}_{N_{2}}$
gauge charges are represented by gauge invariant Wilson operators
\begin{equation}
	\mathsf{P}_{e_{1}00}=\mathcal{N}^{e_{1}00}\exp\left({\rm i}e_{1}\int_{\gamma}A^{1}\right),
\end{equation}
\begin{equation}
	\mathsf{P}_{0e_{2}0}=\mathcal{N}^{0e_{2}0}\exp\left({\rm i}e_{2}\int_{\gamma}A^{2}\right),
\end{equation}
and the pure loop excitations carrying $\mathbb{Z}_{N_{3}}$ gauge
fluxes are represented by 
\begin{equation}
	\mathsf{L}_{00m_{3}}=\mathcal{N}_{00m_{3}}\exp\left({\rm i}m_{3}\int_{\sigma}B^{3}\right),
\end{equation}
where the factors $\mathcal{N}$'s are to be determined. The operator
for pure loop excitation carrying $\mathbb{Z}_{N_{1}}$ gauge fluxes
is 
\begin{align}
	\mathsf{L}_{m_{1}00}= & \mathcal{N}_{m_{1}00}\nonumber\\
	&\times\exp\left[{\rm i}m_{1}\int_{\sigma}B^{1}+\frac{1}{2}\frac{2\pi q}{N_{1}}\left(d^{-1}A^{2}B^{3}+d^{-1}B^{3}A^{2}\right)\right]\nonumber \\
	& \times\delta\left(\int_{\gamma}A^{2}\right)\delta\left(\int_{\sigma}B^{3}\right),
\end{align}
where the Kronecker delta function are $\delta\left(\int_{\gamma}A^{2}\right)=\begin{cases}
	1, & \int_{\gamma}A^{2}=0\mod2\pi\\
	0, & \text{else}
\end{cases}$ and $\delta\left(\int_{\sigma}B^{3}\right)=\begin{cases}
	1, & \int_{\sigma}B^{3}=0\mod2\pi\\
	0, & \text{else}
\end{cases}$ to ensure $d^{-1}A^{2}$ and $d^{-1}B^{3}$ are well-defined \citep{2016arXiv161209298P,PhysRevB.95.035131,PhysRevB.107.165117}.  Similarly, the operator for pure loop carrying $\mathbb{Z}_{N_{2}}$
gauge fluxes is 
\begin{align}
	\mathsf{L}_{0m_{2}0}= & \mathcal{N}_{0m_{2}0}\nonumber\\
	&\times\exp\left[{\rm i}m_{2}\int_{\sigma}B^{2}-\frac{1}{2}\frac{2\pi q}{N_{2}}\left(d^{-1}B^{3}A^{1}+d^{-1}A^{1}B^{3}\right)\right]\nonumber \\
	& \times\delta\left(\int_{\gamma}A^{1}\right)\delta\left(\int_{\sigma}B^{3}\right).
\end{align}
The particle excitation carrying $\mathbb{Z}_{N_{3}}$ gauge charges
is represented by 
\begin{align}
	\mathsf{P}_{00e_{3}}= & \mathcal{N}^{00e_{3}}\exp\left[{\rm i}e_{3}\int_{\gamma}A^{3}+\frac{1}{2}\frac{2\pi q}{N_{3}}\left(d^{-1}A^{1}A^{2}-d^{-1}A^{2}A^{1}\right)\right.\nonumber \\
	& \left.+{\rm i}\frac{e_{3}K_{33}}{N_{3}}\int_{\Sigma}B^{3}\right]\delta\left(\int_{\gamma}A^{1}\right)\delta\left(\int_{\gamma}A^{2}\right),
\end{align}
These Kronecker delta functions can be expanded by summation of some
exponentials, e.g., $\delta\left(\int_{\gamma}A^{1}\right)=\frac{1}{N_{1}}\sum_{k=1}^{N_{1}}\exp\left({\rm i}k\int_{\gamma}A^{1}\right)$
\citep{PhysRevB.95.035131,PhysRevB.107.165117}. As mentioned in previous
section, $\mathsf{P}_{00e_{3}}$ is a particle excitation attached
by a flux string and may be confined due to the tension on string.
$\mathsf{P}_{00e_{3}}$ is deconfined only when the flux on the string
is a multiple of $2\pi$. The minimal $e_{3}$ for deconfined $\mathsf{P}_{00e_{3}}$
is 
\begin{equation}
	e_{3{\rm min}}=\frac{N_{3}}{\gcd\left(\frac{K_{33}}{N_{3}},N_{3}\right)}.
\end{equation}
Notice that the limitation of values of $e_{3}$ influences the period
of $\mathbb{Z}_{N_{3}}$ gauge fluxes. Since a loop excitation is
detected by a particle excitation, we consider the particle-loop braiding
phase of $\mathsf{P}_{00e_{3\min}}$ and $\mathsf{L}_{00m_{3}}$:
\begin{align}
	\Theta_{{\rm PL}}\left(e_{3\min},m_{3}\right) & =\exp\left[-\frac{2\pi e_{3\min}m_{3}}{N_{3}}\right]\nonumber \\
	& =\exp\left[-\frac{2\pi N_{3}m_{3}}{\gcd\left(\frac{K_{33}}{N_{3}},N_{3}\right)}\right].
\end{align}
We immediately see that $m_{3}$ is equivalent with $m_{3}+\gcd\left(\frac{K_{33}}{N_{3}},N_{3}\right)$.
In other words, $m_{3}$ has a period of $\gcd\left(\frac{K_{33}}{N_{3}},N_{3}\right)$.
This is important when we discuss the fusion rules in the following
main text. 

So far we have find out Wilson operators for topological excitation
carrying only one kind of gauge charge or flux. Other excitation with
multiple species of gauge charges or fluxes, e.g., a particle excitation
with different $\mathbb{Z}_{N_{i}}$ gauge charges is defined by
\begin{equation}
	\mathsf{P}_{e_{1}00}\otimes\mathsf{P}_{0e_{2}0}\otimes\mathsf{P}_{00e_{3}}\equiv\mathsf{P}_{e_{1}e_{2}e_{3}},\label{eq:def_fusion_charge}
\end{equation}
or a decorated loop excitation with $\mathbb{Z}_{N_{i}}$ gauge fluxes
and $\mathbb{Z}_{N_{j}}$ gauge charges is defined by 
\begin{equation}
	\mathsf{P}_{e_{1}00}\otimes\mathsf{P}_{0e_{2}0}\otimes\mathsf{P}_{00e_{3}}\otimes\mathsf{L}_{m_{1}00}\otimes\mathsf{L}_{0m_{2}0}\otimes\mathsf{L}_{00m_{3}}\equiv\mathsf{L}_{m_{1}m_{2}m_{3}}^{e_{1}e_{2}e_{3}}.\label{eq:def_fusion_loop}
\end{equation}

Next, we need to determine the factors $\mathcal{N}$ for each operators.
For an illustration, we consider the loop excitation $\mathsf{L}_{100}$
which carries flux of $\mathbb{Z}_{N_{1}}$ gauge subgroup, its Wilson
operator is 
\begin{align}
	\mathsf{L}_{100}= & \mathcal{N}_{100}\exp\left[{\rm i}\int_{\sigma}B^{1}+\frac{1}{2}\frac{2\pi q}{N_{1}}\left(d^{-1}A^{2}B^{3}+d^{-1}B^{3}A^{2}\right)\right]\nonumber \\
	& \times\delta\left(\int_{\gamma}A^{2}\right)\delta\left(\int_{\sigma}B^{3}\right).
\end{align}
Since $\mathsf{L}_{100}$ represent the element $1$ in group $\mathbb{Z}_{N_{1}}$,
according to the $\mathbb{Z}_{N_{1}}$ cyclic structure, it is natural
to require 
\begin{equation}
	\underbrace{\mathsf{L}_{100}\otimes\mathsf{L}_{100}\otimes\cdots\otimes\mathsf{L}_{100}}_{\text{\ensuremath{N_{1}} terms}}=\mathsf{1}+\cdots\label{eq:fusion_N1_L100}
\end{equation}
where ``$\cdots$'' denotes other fusion channels if this fusion
is non-Abelian. Here we have made an assumption: for an excitation
with only kind of charge or flux, fusing it and its anti excitation
would output \emph{one} vacuum. This assumption is reasonable since
a pair of particle and anti particle, or a pair of loop and anti loop,
can be created from vacuum and then be annihilated to vacuum. For
those with multiple kinds of non-Abelian charges or fluxes, fusion
a pair of excitation and anti-excitation may output more than one
vacua \citep{PhysRevB.107.165117}. In path integral, the fusion (\ref{eq:fusion_N1_L100})
is written as 
\begin{widetext}
	\begin{align}
	\left\langle \left(\mathsf{L}_{100}\right)^{\otimes N_{1}}\right\rangle = & \left(\mathcal{N}_{100}\right)^{N_{1}}\left\langle \exp\left[{\rm i}N_{1}\int_{\sigma}B^{1}+\frac{1}{2}\frac{2\pi q}{N_{1}}\left(d^{-1}A^{2}B^{3}+d^{-1}B^{3}A^{2}\right)\right]\right.\nonumber \\
	& \left.\times\delta\left(\int_{\gamma}A^{2}\right)\delta\left(\int_{\sigma}B^{3}\right)\right\rangle \nonumber \\
%	= & \left(\mathcal{N}_{100}\right)^{N_{1}}\left\langle \delta\left(\int_{\gamma}A^{2}\right)\delta\left(\int_{\sigma}B^{3}\right)\right\rangle \nonumber \\
%	= & \left(\mathcal{N}_{100}\right)^{N_{1}}\left\langle \frac{1}{N_{2}}\sum_{e_{2}=1}^{N_{2}}\exp\left({\rm i}e_{2}\int_{\gamma}A^{2}\right)\frac{1}{N_{3}}\sum_{m_{3}=1}^{N_{3}}\exp\left({\rm i}m_{3}\int_{\sigma}B^{3}\right)\right\rangle \nonumber \\
	= & \frac{\left(\mathcal{N}_{100}\right)^{N_{1}}}{N_{2}N_{3}}\left\langle 1+\sum_{e_{2}=1}^{N_{2}-1}\exp\left({\rm i}e_{2}\int_{\gamma}A^{2}\right)+\sum_{m_{3}=1}^{N_{3}-1}\exp\left({\rm i}m_{3}\int_{\sigma}B^{3}\right)\right.\nonumber \\
	& \left.+\sum_{e_{2}=1}^{N_{2}-1}\sum_{m_{3}=1}^{N_{3}-1}\exp\left({\rm i}e_{2}\int_{\gamma}A^{2}+{\rm i}m_{3}\int_{\sigma}B^{3}\right)\right\rangle \label{eq:fusion_output_N1_L100}
\end{align}
\end{widetext}
where we have used 
\begin{align}
	\left\langle \exp\left[{\rm i}N_{1}\int_{\sigma}B^{1}+\frac{1}{2}\frac{2\pi q}{N_{1}}d^{-1}A^{2}B^{3}+d^{-1}B^{3}A^{2}\right]\right\rangle =1
\end{align}
since $B^{1}$ is $\mathbb{Z}_{N_{1}}$ valued. Since the fusion coefficient
of vacuum is $1$, it is required that 
 $
	\frac{\left(\mathcal{N}_{100}\right)^{N_{1}}}{N_{2}N_{3}}=1,
 $
i.e., the factor of Wilson operator for $\mathsf{L}_{100}$ is 
 $
	\mathcal{N}_{100}=\sqrt[N_{1}]{N_{2}N_{3}}.
 $ 
Now we are going to show that the factor $\mathcal{N}_{100}$ is
exactly equal to the quantum dimension of $\mathsf{L}_{100}$. Notice
that the result in Eq. (\ref{eq:fusion_output_N1_L100}) tells us
that the output of fusing $N_{1}$ $\mathsf{L}_{100}$'s is 
\begin{align}
	\left(\mathsf{L}_{100}\right)^{\otimes N_{1}}= & \mathsf{1}\oplus\left(\oplus_{e_{2}=1}^{N_{2}-1}\mathsf{P}_{0e_{2}0}\right)\oplus\left(\oplus_{m_{3}=1}^{N_{3}-1}\mathsf{L}_{00m_{3}}\right)\nonumber \\
	& \oplus\left(\oplus_{e_{2}=1}^{N_{2}-1}\oplus_{m_{3}=1}^{N_{3}-1}\mathsf{L}_{00m_{3}}^{0e_{2}0}\right).\label{eq:fusion_rules_L_100}
\end{align}
It is easy to see that $\mathsf{P}_{0e_{2}0}$ is an Abelian particle
excitation whose quantum dimension is $1$. This is because 
\begin{align}
	\left\langle \left(\mathsf{P}_{010}\right)^{\otimes N_{2}}\right\rangle  & =\left\langle \left(\mathcal{N}^{010}\right)^{N_{2}}\exp\left({\rm i}N_{2}\int_{\gamma}A^{2}\right)\right\rangle ,\nonumber \\
	& =\left\langle \left(\mathcal{N}^{010}\right)^{N_{2}}\cdot1\right\rangle \nonumber \\
	& =\left(\mathcal{N}^{010}\right)^{N_{2}}\cdot\mathsf{1}
\end{align}
where $\mathsf{1}$ denotes the vacuum. Our assumption above requires
$\left(\mathcal{N}^{010}\right)^{N_{2}}=1$ thus $\mathcal{N}^{0e_{2}0}=1,\forall e_{2}\in\mathbb{Z}_{N_{2}}$
. Similarly, we know that $\mathsf{L}_{00m_{3}}$'s and $\mathsf{L}_{00m_{3}}^{0e_{2}0}$'s
are all Abelian excitations. For a fusion rule 
\[
\mathsf{e}_{i}\otimes\mathsf{e}_{k}=\oplus_{m}N_{m}^{ik}\mathsf{e}_{m}
\]
where the quantum dimension of $\mathsf{e}_{i}$ is denoted as $d_{i}$,
there is a relation of these quantum dimensions (the proof can be
found in Appendix \ref{appendix_qdim}): 
\begin{equation}
	d_{i}d_{k}=\sum_{m}N_{m}^{ik}d_{m}.\label{eq:qdim_relation}
\end{equation}
Let the quantum dimension of $\mathsf{L}_{100}$ be $d_{100}$. Applying
Eq. (\ref{eq:qdim_relation}) to fusion rule (\ref{eq:fusion_rules_L_100}),
we have 
\begin{align}
	\left(d_{100}\right)^{N_{1}} & =\sum_{m}N_{m}^{ik}d_{m}=N_{2}N_{3}
\end{align}
thus the quantum dimension of $\mathsf{L}_{100}$ is 
 $
	d_{100}=\sqrt[N_{1}]{N_{2}N_{3}}.
$ 
We can see that the quantum dimension of excitation $\mathsf{L}_{100}$
is just the factor of its Wilson operator. 

Let us go through this line of thinking again: first we write the
Wilson operator of $\mathsf{L}_{100}$ with an unknown factor $\mathcal{N}_{100}$.
At this time we do not know any fusion rules of $\mathsf{L}_{100}$
yet. By demanding $\mathsf{L}_{100}\otimes\mathsf{L}_{\left(N_{1}-1\right)00}=\mathsf{1}+\cdots$
from the $\mathbb{Z}_{N_{i}}$ cyclic structure, we obtain $\mathcal{N}_{100}=\sqrt[N_{1}]{N_{2}N_{3}}$.
Meanwhile, by expanding the Kronecker delta functions, we obtain the
fusion rule (\ref{eq:fusion_rules_L_100}) which tells us the channels
are all Abelian excitations. Since the quantum dimension of Abelian
excitation is $1$, applying Eq. (\ref{eq:qdim_relation}) we find
the quantum dimension of $\mathsf{L}_{100}$ is $d_{100}=\sqrt[N_{1}]{N_{2}N_{3}}$,
same as its Wilson operator's factor. So far, we have seen that for
topological excitation carrying only one species of charge or flux,
its quantum dimension is same as the factor of its Wilson operator.
For topological excitation carrying charges or fluxes from different
$\mathbb{Z}_{N_{i}}$ subgroups, it is defined by fusion of those
with only one kind of charge or flux, see Eqs. (\ref{eq:def_fusion_charge})
and (\ref{eq:def_fusion_loop}). Their quantum dimension can be obtained
by Eq. (\ref{eq:qdim_relation}) and the factor of their Wilson operator
can obtained by path integral calculation according to Eqs. (\ref{eq:def_fusion_charge})
and (\ref{eq:def_fusion_loop}).

We are ready to discuss how the fusion rules of action (\ref{eq:action_AAB+BB})
affected by the $BB$ term. We take an example of $G=\mathbb{Z}_{2}\times\mathbb{Z}_{2}\times\mathbb{Z}_{6}$
and $\frac{K_{33}}{N_{3}}=2$. We will compare the two situations
of $K_{33}=0$ and $\frac{K_{33}}{N_{3}}=2$. The fusion rules of
action (\ref{eq:action_AAB+BB}) without $BB$ term in the
case of $G=\left(\mathbb{Z}_{2}\right)^{3}$ are studied in Ref. \citep{PhysRevB.107.165117}. 

We first take a look at the particle excitation $\mathsf{P}_{00e_{3}}$:
\begin{align}
	\mathsf{P}_{00e_{3}}= & \mathcal{N}^{00e_{3}}\exp\left[{\rm i}e_{3}\int_{\gamma}A^{3}+\frac{1}{2}\frac{2\pi q}{N_{3}}\left(d^{-1}A^{1}A^{2}-d^{-1}A^{2}A^{1}\right)\right.\nonumber \\
	& \left.+{\rm i}\frac{e_{3}K_{33}}{N_{3}}\int_{\Sigma}B^{3}\right]\delta\left(\int_{\gamma}A^{1}\right)\delta\left(\int_{\gamma}A^{2}\right).
\end{align}
As shown in previous discussion, turning on the $\frac{K_{33}}{4\pi}B^{3}B^{3}$
term in action (\ref{eq:action_AAB+BB}) would narrow the choices
of $e_{3}$'s. When $K_{33}=0$, $e_{3}$ takes values from $\mathbb{Z}_{N_{3}}=\mathbb{Z}_{6}$.
When $\frac{K_{33}}{N_{3}}=2$, there exist a minimal value of $e_{3}$,
$e_{3\min}$, and the charges of deconfined $\mathsf{P}_{00e_{3}}$
should satisfy $e_{3}\in e_{3\min}\mathbb{Z}$ where 
\begin{equation}
	e_{3\min}=\frac{N_{3}}{\gcd\left(\frac{K_{33}}{N_{3}},N_{3}\right)}=3.
\end{equation}
In the case of $K_{33}=0$, the charges of $\mathsf{P}_{00e_{3}}$
are labeled by $\mathbb{Z}_{6}$. This $\mathbb{Z}_{6}$ cyclicity
of $e_{3}$ indicates the following fusion rule
\begin{equation}
	\left(\mathsf{P}_{001}\right)^{\otimes6}=\mathsf{1}\oplus\mathsf{P}_{100}\oplus\mathsf{P}_{010}\oplus\mathsf{P}_{110}.
\end{equation}
The quantum dimension of $\mathsf{P}_{001}$ is 
 $	\mathcal{N}^{001}=\sqrt[6]{1+1+1+1}=2^{\frac{1}{3}}.
$ 
In the case of $\frac{K_{33}}{N_{3}}=2$, the charges of deconfined
$\mathsf{P}_{00e_{3}}$ are $3$ and $6$, labeled by $\mathbb{Z}_{\gcd\left(\frac{K_{33}}{N_{3}},N_{3}\right)}=\mathbb{Z}_{2}$.
By definition, $\mathsf{P}_{00e_{3\min}}=\left(\mathsf{P}_{001}\right)^{\otimes3}=\mathsf{P}_{003}$
and its operators is 
\begin{align}
	\mathsf{P}_{00e_{3\min}}= & \mathcal{N}^{00e_{3\min}}\exp\left[{\rm i}e_{3\min}\int_{\gamma}A^{3}\right.\nonumber \\
	& +{\rm i}e_{3\min}\int_{\gamma}\frac{1}{2}\frac{2\pi q}{N_{3}}\left(d^{-1}A^{1}A^{2}-d^{-1}A^{2}A^{1}\right)\nonumber \\
	& \left.+{\rm i}\frac{e_{3\min}K_{33}}{N_{3}}\int_{\Sigma}B^{3}\right]\delta\left(\int_{\gamma}A^{1}\right)\delta\left(\int_{\gamma}A^{2}\right).
\end{align}
Since $\mathsf{P}_{00e_{3}}$ is labeled by $\mathbb{Z}_{2}$ when
$\frac{K_{33}}{N_{3}}=2$, from $\left\langle \mathsf{P}_{00e_{3\min}}\otimes\mathsf{P}_{00e_{3\min}}\right\rangle $
and requiring the coefficient of vacuum is $1$ we have 
\begin{equation}
	\mathsf{P}_{00e_{3\min}}\otimes\mathsf{P}_{00e_{3\min}}=\mathsf{1}\oplus\mathsf{P}_{100}\oplus\mathsf{P}_{010}\oplus\mathsf{P}_{110}.
\end{equation}
Compared to the case of $K_{33}=0$, this is just the fusion rule
of two $\mathsf{P}_{003}$'s. Through this example, we see that one
of the effect of $BB$ term is to confine some particle excitations,
i.e., $\mathsf{P}_{00e_{3}}$ with $e_{3}\neq3\mathbb{Z}$. However,
the fusion rules of deconfined particle excitations are unchanged.
This result can be understood as that the flux attachment due to $BB$
term does not change the particle excitation's internal degrees of
freedom that correspond to fusion. 

Next, we focus on the loop excitation $\mathsf{L}_{00m_{3}}$. As
aforementioned, the $BB$ term makes $m_{3}$ has a smaller
period than $N_{3}$: in the case of $\frac{K_{33}}{N_{3}}=2$, $m_{3}$
is equivalent to $m_{3}+2$. In other words, $m_{3}$ is labeled by
$\mathbb{Z}_{2}$: for $m_{3}\in\left\{ 0,2,4\right\} $, $\mathsf{L}_{00m_{3}}$
is equivalent to the vacuum $\mathsf{1}$; for $m_{3}\in\left\{ 1,3,5\right\} $,
$\mathsf{L}_{00m_{3}}$ is equivalent to $\mathsf{L}_{001}$. The
corresponding fusion rules are:
\begin{align}
	\underbrace{\mathsf{L}_{001}\otimes\mathsf{L}_{001}\otimes\cdots\otimes\mathsf{L}_{001}}_{\text{\ensuremath{N_{3}}=6 terms}}=\mathsf{1}, & K_{33}=0;\\
	\mathsf{L}_{001}\otimes\mathsf{L}_{001}=\mathsf{L}_{002}, & K_{33}=0;\\
	\mathsf{L}_{001}\otimes\mathsf{L}_{001}=\mathsf{1}, & \frac{K_{33}}{N_{3}}=2.
\end{align}

The last example to show is the non-Abelian loop excitation $\mathsf{L}_{100}$:
\begin{align}
	\mathsf{L}_{100}= & \mathcal{N}_{100}\exp\left[{\rm i}\int_{\sigma}B^{1}+\frac{1}{2}\frac{2\pi q}{N_{1}}\left(d^{-1}A^{2}B^{3}+d^{-1}B^{3}A^{2}\right)\right]\nonumber \\
	& \times\delta\left(\int_{\gamma}A^{2}\right)\delta\left(\int_{\sigma}B^{3}\right),
\end{align}
When $K_{33}=0$, these two delta functions can be expanded as
\begin{equation}
	\delta\left(\int_{\gamma}A^{2}\right)=\frac{1}{2}\left[1+\exp\left({\rm i}\int_{\gamma}A^{2}\right)\right],
\end{equation}
\begin{equation}
	\delta\left(\int_{\sigma}B^{3}\right)=\frac{1}{6}\sum_{m_{3}=1}^{6}\exp\left({\rm i}m_{3}\int_{\sigma}B^{3}\right).
\end{equation}
We can calculate the factor $\mathcal{N}_{100}$ from $\left\langle \mathsf{L}_{100}\otimes\mathsf{L}_{100}\right\rangle $:
 $
	\mathcal{N}_{100}=\sqrt{2\times6}=2\sqrt{3}.
 $ As shown in above discussion, $\mathcal{N}_{100}$ is also the quantum
dimension of $\mathsf{L}_{100}$. By setting $\frac{K_{33}}{N_{3}}=2$
we turn on the $BB$ term. Due to the period of $m_{3}$, $m_{3}\simeq m_{3}+\gcd\left(\frac{K_{33}}{N_{3}},N_{3}\right)$,
the expansion of $\delta\left(\int_{\sigma}B^{3}\right)$ actually
becomes (in the sense of correlation with other operators)
\begin{equation}
	\delta\left(\int_{\sigma}B^{3}\right)=\frac{1}{2}\left[1+\exp\left({\rm i}\int_{\sigma}B^{3}\right)\right]
\end{equation}
The factor $\mathcal{N}_{100}$ as well as the quantum dimension of
$\mathsf{L}_{100}$ then becomes
 $	\mathcal{N}_{100}=\sqrt{2\times2}=2.
$

In summary, the influences of $BB$ term on fusion rules are
as follows. First, $BB$ term would confine part of particle
excitations. This in turn makes some loop excitations that used to
distinguishable now become equivalent in the sense of correlation
with other excitations. As in the above example, $\mathsf{L}_{00m_{3}}$
used to be labeled by $\mathbb{Z}_{6}$ but now labeled by $\mathbb{Z}_{2}$
due to the $BB$ term. Consequently, other topological excitations'
quantum dimensions are changed. In the above example, the output of
fusion two $\mathsf{L}_{100}$'s used to be 
\begin{equation}
	\mathsf{L}_{100}\otimes\mathsf{L}_{100}=1\oplus\mathsf{P}_{010}\oplus\left(\oplus_{m_{3}=1}^{6}\mathsf{L}_{00m_{3}}\right)\oplus\left(\oplus_{m_{3}=1}^{6}\mathsf{L}_{00m_{3}}^{010}\right),
\end{equation}
but due to $BB$ term, becomes 
\begin{equation}
	\mathsf{L}_{100}\otimes\mathsf{L}_{100}=1\oplus\mathsf{P}_{010}\oplus\mathsf{L}_{001}\oplus\mathsf{L}_{001}^{010}.
\end{equation}

\section{Conclusion and outlook}\label{section_conclusion}

In this paper, we constructed the topological $BF$ field theory in the presence of both twisted terms (e.g., $AAdA$ and $AAB$) and a $K$-matrix $BB$ term. In this TQFT,  we are allowed to simultaneously explore the self-statistics of particles, particle-loop braiding, multi-loop braiding, Borromean Rings braiding, shrinking rules, and fusion rules, in order to reach a more complete continuum-field-theoretical description of anomaly-free $3$D topological orders. 
We carefully explored the effect of $K$-matrix $BB$ term in two aspects: (i) self-statistics transmutation and (ii) confinement of excitations.  Specially, we illustrated how a general $B   B$ term with a coefficient matrix $K$ alternates the self-statistics of deconfined particle excitations through computing \textit{framed} Wilson loops. We found that the self-statistics of a particle excitation labeled by $\mathbf{l}=\left(e_1,e_2,\cdots,e_n\right)^T$ is given by $\Theta_{\mathbf{l}}=\exp\left(-\pi \mathbf{l}^T \widetilde{K} \mathbf{l}\right)$ where $\widetilde{K}_{ij}=\frac{K_{ij}}{N_i N_j}$ as shown in Eq.~(\ref{eq_general_formula_self}). The expression of this statistical angle is formally very similar to that of a $K$-matrix Chern-Simons theory~\cite{PhysRevB.42.8145,PhysRevB.46.2290}.
We also examined in what situation, respectively, trivial fermions (fermionic trivial particles) and emergent fermions (fermionic particles that carry nontrivial gauge charges) are
possible and how they influence   braiding statistics and fusion rules studied in Ref.~\cite{PhysRevB.107.165117}.

If $3$-loop braiding and/or BR braiding are considered,
the loops are allowed to carry gauge fluxes from  different gauge subgroups.
We found that for those gauge subgroups whose gauge fluxes take part in $3$-loop
braiding~\cite{wang_levin1} or BR braiding~\cite{yp18prl}, their gauge charges can only be carried by
bosonic particle excitations. This result is obtained from the incompatibility between $AAdA$ twisted term and $BB$ term within our framework of continuum field theory. Physically this can be interpreted as that these two topological terms have different microscopic origin (see Sec.~\ref{subsec_condensation_picture}). For example, when $G=\mathbb{Z}_{N_{1}}\times\mathbb{Z}_{N_{2}}$
and $3$-loop braidings are considered, all particle excitations are bosonic, i.e., emergent fermions are forbidden. Furthermore, we take BR topological
order as an example to see how emergent fermion influences its fusion rules. 

For the future directions, it is interesting to write all compatibility conditions proposed in Ref.~\cite{zhang2021compatible} and the present work in a more symbolical way and compare the continuum-field-theoretical analysis and the mathematics of higher category. Due to the general belief on the bulk-boundary correspondence, we may  examine the $(2+1)$D boundary theory by placing TQFTs on an open manifold, in order to understand compatibility from boundary.

\acknowledgments
We thank M. Cheng for the enlightening discussions. Z.F.Z \& P.Y. were supported in part by NSFC Grant No.~12074438, 
Guangdong Basic and Applied Basic Research Foundation under Grant
No.~2020B1515120100, and the Open Project of Guangdong
Provincial Key Laboratory of Magnetoelectric Physics and Devices under
Grant No.~2022B1212010008. Z.F.Z \& P.Y.  were also supported in part by the Fundamental Research Funds for the Central Universities, and the Research Funds of Sun Yat-sen University. Q.R.W. was supported in part by NSFC Grant No.~12274250.

%\bibliography{braid}

\appendix

\section{The relation of quantum dimensions in a fusion rule}\label{appendix_qdim}

Let the quantum dimension of $\mathsf{e}_{i}$ be $d_{i}$. Now we
prove that for 
\begin{equation}
	\mathsf{e}_{i}\otimes\mathsf{e}_{k}=\oplus_{m}N_{m}^{ik}\mathsf{e}_{m}
\end{equation}
one can know 
\begin{equation}
	d_{i}d_{k}=\sum_{m}N_{m}^{ik}d_{m}.
\end{equation}
From the associativity of fusion rules, we have 
\begin{align}
	& (\mathsf{e}_{i}\otimes\mathsf{e}_{j})\otimes\mathsf{e}_{k}=\mathsf{e}_{i}\otimes(\mathsf{e}_{j}\otimes\mathsf{e}_{k}).
\end{align}
The left hand side can be written as $\oplus_{m}N_{m}^{ij}\mathsf{e}_{m}\otimes\mathsf{e}_{k}=\oplus_{m}N_{m}^{ij}\oplus_{l}N_{l}^{mk}\mathsf{e}_{l}$
and the right hand side can be written as $\mathsf{e}_{j}\otimes\left(\oplus_{m}N_{m}^{ik}\mathsf{e}_{m}\right)=\oplus_{m}N_{m}^{ik}\oplus_{l}N_{l}^{jm}\mathsf{e}_{l}$.
The fusion coefficients $N_{k}^{ij}$'s can form a matrix $N_{i}$
with $\left(N_{i}\right)_{kj}=N_{k}^{ij}$. Therefore we have 
\begin{equation}
	\oplus_{m}N_{m}^{ij}N_{l}^{mk}=\oplus_{m}N_{m}^{ik}N_{l}^{jm}.
\end{equation}
Notice that $\oplus_{m}N_{m}^{ij}N_{l}^{mk}=\sum_{m}N_{m}^{ij}N_{l}^{mk}=\sum_{m}\left(N_{i}\right)_{mj}\left(N_{k}\right)_{lm}=\left(N_{i}N_{k}\right)_{lj}$
and $\oplus_{m}N_{m}^{ik}N_{l}^{im}=\sum_{m}N_{m}^{ik}N_{l}^{jm}=\sum_{m}N_{m}^{ik}\left(N_{m}\right)_{lj}$,
where we have used $N_{c}^{ab}=N_{c}^{ba}$. So we have a relation
between matrices:
\begin{equation}
	N_{i}N_{k}=\sum_{m}N_{m}^{ik}N_{m}.
\end{equation}
Since $N_{i}$'s are commutative, their largest eigenvalues $d_{i}$'s,
i.e., quantum dimensions of corresponding topological excitations,
satisfy 
\begin{equation}
	d_{i}d_{k}=\sum_{m}N_{m}^{ik}d_{m}.
\end{equation}

\end{document}